\documentclass[twocolumn,showpacs,amsmath,amssymb,prd]{revtex4}
\usepackage{graphicx}
\newcommand{\beq}{\begin{equation}}
\newcommand{\eeq}{\end{equation}}
\newcommand{\beqn}{\begin{eqnarray}}
\newcommand{\eeqn}{\end{eqnarray}}

\newcommand{\ve}[1]{\mbox{\boldmath $#1$}}

\newcommand{\Thydro}[2]{T_{\mathrm{(hydro)}}^{#1 #2}}

\newcommand{\Tem}[2]{T_{\mathrm{(em)}}^{#1 #2}}
\newcommand{\Tmhd}[2]{T_{(\mathrm{mhd})}^{#1 #2}}
\newcommand{\Tmhdlow}[2]{T_{(\mathrm{mhd})#2}^{#1}}
\newcommand{\Ttotal}[2]{T^{#1 #2}}
\newcommand{\Ttotallow}[2]{T^{#1}{}_{#2}}
\newcommand{\Trad}[2]{R^{#1 #2}}
\newcommand{\Tradlow}[2]{R^{#1}{}_{#2}}
\newcommand{\Erad}{E}
\newcommand{\Frad}{F}
\newcommand{\Prad}{\mathcal{P}}
\newcommand{\taurad}{\bar{\tau}}
\newcommand{\Srad}[1]{\bar{S}_{#1}}
\newcommand{\srad}{\bar{s}}
\newcommand{\rhostar}{\rho_*}
\newcommand{\taumhd}{\tilde{\tau}}
\newcommand{\Smhd}[1]{\tilde{S}_{#1}}

\newcommand{\sg}{\sqrt{\gamma}\,}
\newcommand{\proj}[2]{h^{#1 #2}}
\newcommand{\projlow}[2]{h^{#1}{}_{#2}}

\usepackage{dcolumn}
\usepackage{bm}
\usepackage{epsf}

\begin{document}
\bibliographystyle{apsrev}
\title{Relativistic Radiation Magnetohydrodynamics in Dynamical
 Spacetimes: Numerical Methods and Tests}

\author{Brian D. Farris}

\author{Tsz Ka Li}

\author{Yuk Tung Liu}

\author{Stuart L.\ Shapiro}
\altaffiliation{Also at Department of Astronomy \& NCSA, University of Illinois
at Urbana-Champaign, Urbana, IL 61801}

\affiliation{Department of Physics, University of Illinois at
Urbana-Champaign, Urbana, IL~61801}

\begin{abstract}
Many systems of current interest in relativistic astrophysics require 
a knowledge of radiative transfer in a magnetized gas flowing 
in a strongly-curved, {\it dynamical} spacetime.  
Such systems include coalescing compact binaries containing neutron stars 
or white dwarfs, disks around merging black holes, core collapse supernovae, 
collapsars, and gamma-ray burst sources. To model
these phenomena, all of which involve general relativity,
radiation (photon and/or neutrino), and magnetohydrodynamics, we have developed a 
general relativistic code capable of evolving MHD fluids 
and radiation in dynamical
spacetimes. Our code solves the coupled Einstein-Maxwell-MHD-Radiation system
of equations both in axisymmetry and in full 3+1 dimensions.  We
evolve the metric by integrating the
BSSN equations, and use a conservative, high-resolution 
shock-capturing scheme to evolve both the MHD and radiation 
moment equations. In this paper, we implement our scheme for 
optically thick gases and grey-body opacities.  
Our code gives accurate results in a suite of tests involving radiating shocks
and nonlinear waves propagating in Minkowski spacetime. In addition, 
to test our code's ability to evolve the relativistic radiation-MHD 
equations in strong-field dynamical
spacetimes, we study ``thermal Oppenheimer-Snyder collapse'' to 
a black hole, and find
good agreement between analytic and numerical solutions.
\end{abstract}


\maketitle

\section{Introduction}
\label{intro}
Many relativistic systems of current astrophysical interest
are characterized by the dynamical coupling of strong-field
gravitation, high magnetic fields and intense
radiation (where the latter may
be photons or neutrinos).  Quasars, active galactic
nuclei (AGNs), galactic ``micro-quasars'', core-collapse
supernovae, collapsars, gamma-ray burst sources (GRBs), merging neutron star
binaries (NSNSs), merging black hole-neutron star binaries
(BHNSs), and merging neutron star-white dwarf binaries (NSWDs)
are all examples of such systems.
Developing robust computational methods that
can treat simultaneously the different dynamical
phenomena that govern these systems
is necessary in order to simulate their physical behavior 
reliably and identify their observational signatures. 

Many of the systems listed above involve compact objects,
such as black holes and neutron stars. Hence general
relativity is required to describe their dynamical evolution accurately.
Both observations and theory strongly suggest that
magnetic fields play an important role in many of these systems. For example, 
magnetic fields are crucial in launching jets
from black holes in AGNs and GRBs 
(see, e.g., \cite{bbr84,piran05}), driving
accretion onto black holes in disks (see, e.g., 
\cite{bh98,mrf06}), and inducing `delayed'
collapse in hypermassive neutron stars that may form following 
NSNS mergers~\cite{bss00,s00,dlsss06a,dlsss06b}.
Radiation, apart from
its role as an observational tracer and diagnostic probe,
also can play an important dynamical role in many relativistic
systems.  
For example, the role of neutrino transport may be essential
to understanding core-collapse supernovae (see, 
e.g.~\cite{jlmmm07,bdol07,mbbhm07}). 
As a second example, consider that 
the interior pressure of supermassive stars and massive
Population~III stars is dominated by thermal
radiation pressure. These objects may collapse in the early
universe to form the seeds
of the supermassive black holes that reside in the centers
of many, and perhaps most, galaxies~\cite{richstone,ho99}. Radiation thus
plays a crucial role in determining the onset and
dynamics of the collapse of these stars
and the masses and spins of the black holes that are 
formed~\cite{bs99,stu04,shapiro05}.
As a final example, accretion onto compact objects
leading to outgoing radiation near and above the Eddington value
is controlled by the competition between inward gravitational forces and
outward radiation pressure forces. All of these systems need to be
handled in a computational scheme designed to probe
these physical phenomena self-consistently, simultaneously 
accounting for radiation, magnetic fields and relativistic gravitation.

We have developed previously a robust numerical scheme in 3+1 dimensions 
that simultaneously
evolves the Einstein equations of general relativity 
for the gravitational field (metric),
the equations of relativistic magnetohydrodynamics (MHD)
for the matter, and Maxwell's equations for a magnetic
field~\cite{mhd_code_paper}. Our approach is based on the BSSN
(Baumgarte-Shapiro-Shibata-Nakamura) formalism to treat the
gravitational field~\cite{bssn_shibata,bssn_stu},
a high-resolution, shock-capturing (HRSC) scheme to handle
the fluid and a constrained-transport scheme to treat
 magnetic induction~\cite{t00}. Our resulting
GRMHD code has been subjected to a rigorous suite of numerical
tests to check and calibrate its validity~\cite{mhd_code_paper}.
We have applied our code to explore
a number of dynamical scenarios, including the
collapse of magnetized, differentially rotating,
hypermassive neutron stars to black holes~\cite{dlsss06a,dlsss06b},
the collapse of rotating stellar cores to neutron stars~\cite{slss06},
the collapse of rotating, supermassive stars and massive
Pop~III stars to black holes~\cite{lss07}, and the merger of
binary black holes~\cite{junk-ID} and binary black hole-neutron 
stars~\cite{eflstb07}.
The purpose of this paper is to present a generalization of
our current GRMHD scheme that accounts for the presence of
radiation (photon or neutrino).

Our approach for handling the radiation
follows in the long tradition of formalisms
designed to treat radiation transport in the framework of
general relativity. However, we have developed a new version
specifically designed to fit neatly onto our existing $3+1$ GRMHD scheme.
The general relativistic radiative transfer equation
has been derived in full detail
by Lindquist in 1966~\cite{lindquist66}.
His treatment has been followed by numerous adaptations and
implementations in various approximations.
For example, Thorne has
derived a set of radiation moment equations to arbitrary order
by the technique of projected symmetric trace-free (PSTF) 
tensors~\cite{thorne_pstf}. So far, most
GR radiation hydrodynamics calculations 
(e.g.~\cite{schmid-burgk78,schinder88,stu_rad_1,stu_rad_2,
sb89,mm89,ztne96,lmmbct04}), including those based on the PSTF scheme 
(e.g.~\cite{tfz81,RezM94,zampieri,tzzn96,RezM96,bzs00}), 
have been implemented in spherical symmetry only. 
Once spherical symmetry is broken, most radiation schemes become 
quite difficult to
implement, given the large number of phase space degrees of freedom
that need to be tracked for the radiation field.

In this paper, we formulate
the radiation transport equations in the framework of
our $3+1$ GRMHD scheme, which operates
without any restrictions regarding the
spatial symmetry of the system.
However, our implementation focuses on the {\it optically thick} limit for
the radiation field,
which simplifies the analysis by allowing us to
assume that the radiation field in the comoving frame of the fluid
is nearly isotropic.
Our emphasis is geared to treating systems in which
the radiation has a strong dynamical influence on the
matter flow and, in some cases, on the
spacetime geometry itself. We are less concerned in this
initial treatment with the radiation that escapes from the matter surface,
or with the radiation spectrum measured by a distant observer.
It is in the interiors of collapsing stars,
neutron stars in merging compact binaries, and
dense accretion disks orbiting black holes
where the dynamical influence of the radiation field is
likely to play its most significant role. In these interior regions
the optically thick assumption should be quite reliable
in many cases. In the implementation
presented here we also adopt a {\it grey-body opacity law}, which, though
simple, suffices to illustrate our method. However, the formalism
makes no assumptions regarding the spatial symmetry of the matter source,
radiation field, or spacetime.

We present two sets of tests to check
our new radiation GRMHD code. The first set of tests involves
radiation shocks and nonlinear
waves propagating in a fixed Minkowski spacetime.
The second set of tests is  the ``thermal Oppenheimer-Snyder
collapse'' problem originally proposed and solved
by Shapiro~\cite{stu_rad_1,stu_rad_2}, wherein radiation propagates in a
spherical spacetime that, though simple, is highly dynamical 
and characterized by a strong
gravitational field (i.e. one in which a black hole forms).
In both sets of tests, we compare our numerical results with
analytic solutions and perform convergence tests.

The structure of the paper is as follows:
In Sec.~\ref{formalism},
we formulate the system of coupled
Einstein-Maxwell-MHD-radiation
equations in $3+1$ form, with the Maxwell,
MHD and radiation equations written in conservative form.
In Sec.~\ref{implementation},
we describe techniques for evolving this system of equations.
In Sec.~\ref{tests}, we present the new
code tests and their results.
Finally, we summarize our results in Sec.~\ref{conclusions}.

\section{Formalism}
\label{formalism}
Throughout this paper, Latin indices denote spatial components
(1-3) and Greek indices denote spacetime components (0-3). 
We adopt geometrized units, so that $G = c = 1$.

\subsection{Evolution of gravitational fields}

We write the spacetime metric in the standard 3+1 form:
\begin{equation}
ds^2 = -\alpha^2 dt^2 + \gamma_{ij}(dx^i+\beta^idt)(dx^j+\beta^jdt),
\end{equation}
where $\alpha$, $\beta^i$, and $\gamma_{ij}$ are the lapse, shift, and
spatial metric, respectively.  The extrinsic curvature $K_{ij}$ is
defined by
\begin{equation}
\label{Kij}
(\partial_t - {\mathcal{L}}_{\beta})\gamma_{ij} = -2\alpha K_{ij},
\end{equation}
where ${\mathcal{L}}_{\beta}$ is the Lie derivative with respect to
$\beta^i$. The evolution of $\gamma_{ij}$ and $K_{ij}$ is governed 
by the Einstein equation $G_{\mu \nu} = 8\pi T_{\mu \nu}$, where 
$G_{\mu \nu}$ is the Einstein tensor and $T_{\mu \nu}$ is the stress-energy 
tensor. 

We evolve $\gamma_{ij}$ and $K_{ij}$
using the BSSN 
formulation~\cite{bssn_shibata,bssn_stu}.  The fundamental variables for
BSSN evolution are
\begin{eqnarray}
  \phi &\equiv& \frac{1}{12}\ln[\det(\gamma_{ij})]\ , \\
  \tilde\gamma_{ij} &\equiv& e^{-4\phi}\gamma_{ij}\ , \\
  K &\equiv& \gamma^{ij}K_{ij}\ , \\
  \tilde A_{ij} &\equiv& e^{-4\phi}(K_{ij} - \frac{1}{3}\gamma_{ij}K)\ , \\
  \tilde\Gamma^i &\equiv& -\tilde\gamma^{ij}{}_{,j}\ .
\end{eqnarray}
The Einstein equation $G_{\mu \nu} = 8\pi T_{\mu \nu}$ gives rise 
to the evolution equations and constraint equations for these 
fields, which are summarized in~\cite{bssn_stu}.  
In this paper, we use the same field evolution equations as Eqs.~(11)--(15) 
of~\cite{dmsb03}:
\begin{eqnarray}
\label{evolve_gamma}  
(\partial_t - {\mathcal{L}}_{\beta})\tilde\gamma_{ij}
                 &=& -2\alpha\tilde A_{ij}, \\
\label{evolve_phi}
(\partial_t - {\mathcal{L}}_{\beta})\phi
                 &=& -{1\over 6}\alpha K, \\
\label{evolve_K}
(\partial_t - {\mathcal{L}}_{\beta})K
                 &=& -\gamma^{ij}D_jD_i\alpha + {1\over 3}\alpha K^2 \\
                 & & + \alpha \tilde A_{ij}\tilde A^{ij}
                     + 4\pi\alpha (\rho + S), \nonumber \\
\label{evolve_A}
(\partial_t - {\mathcal{L}}_{\beta})\tilde A_{ij}
                 &=& e^{-4\phi}(-D_iD_j\alpha
                     + \alpha(R_{ij}-8\pi S_{ij}))^{TF} \nonumber \\
                 & & + \alpha(K\tilde A_{ij} - 2\tilde A_{il}\tilde
A^l{}_j),
\end{eqnarray}
and
\begin{eqnarray}
\label{evolve_Gamma} 
\partial_t\tilde\Gamma^i &=& \partial_j(2\alpha\tilde A^{ij}
                 + {\mathcal{L}}_{\beta}\tilde\gamma^{ij}) \nonumber \\
         &=& \tilde\gamma^{jk}\beta^i{}_{,jk}
                 + {1\over 3}\tilde\gamma^{ij}\beta^k{}_{,kj}
  - \tilde\Gamma^j\beta^i{}_{,j} \\
  & &+{2\over 3}\tilde\Gamma^i\beta^j{}_{,j}
     + \beta^j\tilde\Gamma^i{}_{,j} - 2\tilde A^{ij}\partial_j\alpha
         \nonumber \\
  & &- 2\alpha\left({2\over 3}\tilde\gamma^{ij}K_{,j} - 6\tilde
A^{ij}\phi_{,j}
     - \tilde\Gamma^i{}_{jk}\tilde A^{jk} + 8\pi\tilde\gamma^{ij}S_j
         \right),
         \nonumber
\end{eqnarray}
where $D$ denotes covariant derivative operator associated with
$\gamma_{ij}$, and $TF$ denotes the trace-free part of a tensor.
The constraint equations, expressed in terms of the BSSN variables, are 
\begin{eqnarray}
\label{Hamiltonian_BSSN}
  0 = \mathcal{H} &=&
                \tilde\gamma^{ij}\tilde D_i\tilde D_j e^{\phi}
                - {e^{\phi} \over 8}\tilde R  \\
                & & + {e^{5\phi}\over 8}\tilde A_{ij}\tilde A^{ij}
                    - {e^{5\phi}\over 12}K^2 + 2\pi e^{5\phi}\rho,
        \nonumber \\
\label{momentum_BSSN}
  0 = {\mathcal{M}}^i &=&
  \tilde D_j(e^{6\phi}\tilde A^{ji})- {2\over 3}e^{6\phi}\tilde D^i K
  - 8\pi e^{6\phi}S^i,
\end{eqnarray}
where $\tilde D$ denotes covariant derivative operator associated with 
$\tilde{\gamma}_{ij}$.
The matter-energy source terms are given by 
\begin{eqnarray}
  \rho &=& n_{\alpha}n_{\beta}\Ttotal{\alpha}{\beta}\ , \nonumber \\
  S_i  &=& -\gamma_{i\alpha}n_{\beta}\Ttotal{\alpha}{\beta}\ , \label{source_def} \\
  S_{i j} &=& \gamma_{i\alpha}\gamma_{j\beta}\Ttotal{\alpha}{\beta}\ ,\nonumber\\
  S &=& \gamma^{ij} S_{ij}\ .\nonumber
\end{eqnarray}
Here $n^{\alpha} = (\alpha^{-1},-\alpha^{-1} \beta^i)$ is the time-like
unit vector normal to the $t=$ constant time slices. In this paper
$\Ttotal{\alpha}{\beta}$ contains three components:
\begin{equation}
\Ttotal{\alpha}{\beta}=\Thydro{\alpha}{\beta}+\Tem{\alpha}{\beta}+\Trad{\alpha}{\beta}
\end{equation}
where $\Thydro{\alpha}{\beta}$, $\Tem{\alpha}{\beta}$ and
$\Trad{\alpha}{\beta}$ are the stress-energy tensor for the hydrodynamic
matter field, (large scale) electrodynamic field and the radiation field, 
respectively. Hence all components here contribute to the BSSN source terms in
Eq.~(\ref{source_def}). 

In order to evolve the 3+1 Einstein equations forward in time, one
must choose lapse $\alpha$ and shift $\beta^i$ functions, which
specify how the spacetime is foliated. 
The lapse and shift must be chosen in such a way that the total
system of evolution equations is stable.  
In the past few years, we have experimented with several 
gauge conditions. We find that, in general, the most useful gauge choices are 
the hyperbolic driver conditions~\cite{abpst01,hydro_excision}, and 
the puncture gauge conditions (see e.g.~\cite{rit06,goddard06}). 
In this paper, we use the hyperbolic driver conditions as 
in~\cite{hydro_excision} when evolving a dynamical spacetime:
\begin{eqnarray}
\label{hb_lapse_nok3}
\nonumber
\partial_t \alpha &=& \alpha {\cal A} \\
\partial_t {\cal A} &=& -a_1(\alpha\partial_tK + a_2{\cal A} +
a_3 e^{-4\phi}\alpha K)\ . \\ 
\label{hb_shift}
\partial^2_t\beta^i &=& b_1\alpha\partial_t\tilde\Gamma^i
 - b_2\partial_t\beta^i\ ,
\end{eqnarray}
where $a_1$, $a_2$, $a_3$, $b_1$, and $b_2$ are freely specifiable
constants.

\subsection{Evolution of radiation fields}
\label{sec:radiation-fields}

\subsubsection{Radiation fields}

The equations governing the dynamics of the radiation can be expressed
as
\begin{equation}
\label{rad_evolution}
\Trad{\alpha}{\beta}{}_{;\beta} = - G^{\alpha}
\end{equation}
where $\Trad{\alpha}{\beta}$ is the radiation stress-energy tensor, and
$G^{\alpha}$ is the radiation four-force density which describes the
interaction of the matter with the radiation 
\cite{mihalas,stu_rad_2}. The radiation stress-energy
tensor $\Trad{\alpha}{\beta}$ is defined as 
\begin{equation} \Trad{\alpha}{\beta} = \int d \nu d \Omega I_{\nu}
N^{\alpha}N^{\beta}
\end{equation}
where $\nu$ is the frequency, $I_{\nu} = I(x^{\alpha};N^i,\nu)$ is the specific
intensity of radiation at $x^{\alpha}$ moving in the direction 
$N^{\alpha} \equiv
p^{\alpha}/h \nu$, $p^{\alpha}$ is the photon 4-momentum, $h$ 
is the Planck constant, and $d\Omega$ is the 
differential solid angle. Here $\nu$, $I_{\nu}$ and $d\Omega$ are all
measured in the local Lorentz frame of a fiducial observer with
4-velocity $u^{\alpha}_{\mathrm{(fid)}}$, i.e. $h \nu = - p_{\alpha}u^{\alpha}_{\mathrm{(fid)}}$.  The integral is evaluated over all frequency 
and solid angles.

We now choose our fiducial observer to be comoving with the fluid.  In the comoving frame of the fluid the radiation stress-energy tensor
$\Trad{\alpha}{\beta}$ takes the form 
\begin{equation}
\label{comoving}
\Trad{\hat{\alpha}}{\hat{\beta}} = \left[ 
\begin{array}{l l l l}
\Erad & \Frad^{\hat{x}} & \Frad^{\hat{y}} & \Frad^{\hat{z}}\\
\Frad^{\hat{x}} & \Prad^{\hat{x} \hat{x}} & \Prad^{\hat{x}\hat{y}} & \Prad^{\hat{x}\hat{z}}\\
\Frad^{\hat{y}} & \Prad^{\hat{y} \hat{x}} & \Prad^{\hat{y}\hat{y}} & \Prad^{\hat{y}\hat{z}}\\
\Frad^{\hat{z}} & \Prad^{\hat{z} \hat{x}} & \Prad^{\hat{z}\hat{y}} & \Prad^{\hat{z}\hat{z}}\\
\end{array}
\right]
\end{equation}
where
\begin{equation}
\Erad = \int d \nu d \Omega I_{\nu}
\end{equation}
is the comoving radiation energy density,
\begin{equation}
\Frad^{\hat{\imath}} = \int d \nu d \Omega I_{\nu}N^{\hat{\imath}}
\end{equation}
is the comoving radiation flux, and
\begin{equation}
\Prad^{\hat{\imath} \hat{\jmath}} = \int d \nu d \Omega I_{\nu}N^{\hat{\imath}}N^{\hat{\jmath}}
\end{equation}
is the comoving radiation stress tensor.

We are interested in the optically thick regime, in which the
radiation is very nearly isotropic in the comoving frame of the
fluid. In the limit of {\it strict} isotropy, independent of the
propagation direction $N^{\hat{\imath}}$, the intensity is $I_\nu=I(x^\alpha;\nu)$. Using this fact
and the expression of $N^\alpha$ in the comoving frame 
\begin{equation}
N^{\hat{\alpha}} = (1,N^{\hat{\imath}})= (1,\mathrm{sin} \theta \mathrm{cos} \varphi, \mathrm{sin}
\theta \mathrm{sin} \varphi,\mathrm{cos} \theta),
\end{equation}
one can show that $\Frad^{\hat{\imath}}=0$ and
$\Prad^{\hat{\imath} \hat{\jmath}} = \frac{1}{3}\delta^{\hat{\imath} \hat{\jmath}}
\Erad\equiv\delta^{\hat{\imath} \hat{\jmath}}\Prad$, where $\Prad$ is the
radiation pressure, $\theta$ is the polar angle measured from 
the $\hat{z}$-axis, 
and $\varphi$ is the azimuthal angle 
(i.e.\ $\tan \varphi = N^{\hat{y}}/N^{\hat{x}}$) . 
Henceforth, we include the effect of 
a small anisotropy by allowing a small non-zero radiation flux 
$\Frad^{\hat{\imath}}$, but we retain the closure relation 
$\Prad=\Erad/3$.  That is, we adopt an Eddington factor equal to $1/3$.

The radiation stress-energy tensor $\Trad{\alpha}{\beta}$ can be
written in covariant form as 
\begin{equation}
\Trad{\alpha}{\beta} = \Erad u^{\alpha}u^{\beta} +
\Frad^{\alpha}  u^{\beta} +u^{\alpha}  \Frad^{\beta} +
\Prad\proj{\alpha}{\beta} \ ,
\label{R_def}
\end{equation}
where $u^{\alpha}$ is the fluid 4-velocity.  This expression reduces to the same form as Eq.~(\ref{comoving}) in the comoving
frame. Here we have introduced the projection tensor, $\proj{\alpha}{\beta}$, 
defined as
\begin{equation}
\proj{\alpha}{\beta} = g^{\alpha \beta} + u^{\alpha}u^{\beta} \ ,
\end{equation}
and the radiation flux four-vector defined as
\begin{equation}
\Frad^{\alpha} = \projlow{\alpha}{\beta} \int  d \nu d \Omega I_{\nu}
N^{\beta}.
\end{equation}
Note that with this definition, the flux satisfies
\begin{equation} 
\Frad^{\alpha}u_{\alpha} = 0.
\end{equation}
Following \cite{stu_rad_2}, the radiation four-force density is given by 
\begin{equation}
G^{\alpha}=\int d \nu d \Omega ( \chi_{\nu} I_{\nu} - \eta_{\nu})
N^{\alpha} \ ,
\end{equation}
where $\chi_{\nu}=\chi^a_\nu+\chi^s_\nu$ is the total opacity (the
superscript $a$ and $s$ denote the absorption and scattering
opacities respectively) and $\eta_{\nu} = \eta_{\nu}^a + \eta_{\nu}^s$ is the total
emissivity. By assuming isotropic and coherent scattering, and that
the thermal emissivity $\eta_{\nu}^a$ and absorption coefficient
$\chi^a_\nu$ are related by Kirchhoff's law $\eta_\nu^a=\chi^a_\nu
B_\nu$, we can write, in the fluid comoving frame, 
\begin{eqnarray}
G^{\hat{0}} &=& \int d \nu d \Omega (\chi^a_{\nu} I_{\nu} -
\eta^a_\nu) = \int d \nu d \Omega \chi^a_{\nu} (I_{\nu} - B_{\nu}) \ ,\cr
G^{\hat{\imath}} &=& \int d \nu d \Omega (\chi^a_{\nu} + \chi^s_{\nu})I_{\nu}
N^{\hat{\imath}} \ ,
\end{eqnarray}
where $B_{\nu}$ is the intensity in thermal equilibrium (e.g. the Planck
function for photons, the analogous Fermi-Dirac function for
neutrinos, etc.). We further assume a grey-body form for all
opacities, $\chi_\nu = \kappa \rho_0$, where $\kappa$ is a frequency
independent opacity, and $\rho_0$ is the rest mass-energy
density. Then we may write \cite{stu_rad_2}
\begin{eqnarray}
G^{\hat{0}} &=&  \rho_0 \kappa^a (\Erad - 4 \pi B) \ ,\cr
G^{\hat{\imath}} &=&  \rho_0(\kappa^a + \kappa^s) \Frad^{\hat{\imath}} \ .
\end{eqnarray}
It is straightforward to express $G^\alpha$ in covariant form as
\begin{equation}
G^{\alpha} =  \rho_0 \kappa^a (\Erad - 4 \pi B)u^{\alpha}+
 \rho_0(\kappa^a + \kappa^s) \Frad^{\alpha} \ .
\label{eq:Ga}
\end{equation}
Note that the frequency integrated equilibrium intensity $B(T)$ can be written as
\begin{equation}
4 \pi B = a_R T^4 \ ,
\label{B_def}
\end{equation}
where $T$ is the temperature of the fluid, and $a_R$ is a constant depending
on the type of radiation: for thermal photons it equals the usual radiation
constant $a$; for each flavor of nondegenerate thermal neutrino or antineutrino
(chemical potentials = 0) it is $(7/16)a$; lumping the contributions of all
neutrinos and antineutrinos together, it is $(7\mathcal{N_\nu}/8)a$, 
where $\mathcal{N_\nu}$ is the number of neutrino flavors which
contribute to thermal radiation.

We emphasize here that our method allows for 
situations in which the gas may be out of thermal equilibrium with
the radiation $(E \ne 4\pi B)$.  Our formalism is equivalent to
keeping the first two
radiative moment equations, and using an Eddington factor to close the
set.  Our choice of $\Prad =
1/3 \Erad$ serves as the necessary closure relation for these
equations.  We demonstrate in Appendix~\ref{diff_approx_appendix} 
that our formalism, while more general, reduces to the relativistic
diffusion approximation in a simplifying limit. In this limit, 
the diffusion approximation transforms the radiation moment 
evolution equations from a hyperbolic to a parabolic (i.e.\ diffusion) form, 
which does not have the same causal structure as the original system
of equations.  Moreover, the parabolic form is not suitable for
implementing the conservative HRSC scheme used to 
integrate the combined MHD-radiation equations (see
Sec.~\ref{implementation}).  In any case, we do not adopt the
diffusion approximation here, but treat the original set without simplification.

\subsubsection{Radiation evolution}

We can decompose the radiation evolution equations given by 
(\ref{rad_evolution}) in a manner analogous to the way we
decompose the MHD evolution equations (e.g., see Sec.~IIC 
in~\cite{mhd_code_paper} and Sec.~\ref{mhd_evolution_section} 
below). The resulting equations are therefore cast in conservative form, 
as are the MHD evolution equations. 
Taking the scalar product of Eq.~(\ref{rad_evolution}) with 
$n_{\alpha}$ on both sides gives the energy equation
\begin{eqnarray}
\partial_t \taurad+\partial_i(\alpha^2\sg R^{0i})&=&\srad-(\alpha^2\sg)G^0 \ ,
\label{rad_energy}
\end{eqnarray}
where the radiation energy density variable $\taurad$ is
defined as 
\begin{eqnarray}
\taurad&=&(\alpha^2\sg)R^{00}\cr
&=&\sg (\alpha u^0)^2 \frac{4}{3}\Erad + 2 \sg\alpha^2 u^0 \Frad^0 -
\sg \frac{1}{3}\Erad \ ,
\label{taurad_def}
\end{eqnarray}
and the source term $\srad$ is
\begin{eqnarray}
\srad&=&-\alpha\sg R^{\mu\nu}\nabla_{\nu} n_{\mu}\cr
&=&\alpha \sg \left[ (\Trad{0}{0} \beta^i \beta^j + 2 \Trad{0}{i} \beta^j
+ \Trad{i}{j})K_{ij}\right. \cr
&& \left. - (\Trad{0}{0} \beta^i + \Trad{0}{i})\partial_i
\alpha \right] \ .
\end{eqnarray}
Here $\gamma=e^{12 \phi}$ denotes the determinant of the spatial metric 
$\gamma_{ij}$. 
The spatial components of Eq.~(\ref{rad_evolution}) give
the momentum equation,
\begin{eqnarray}
\label{rad_momentum}
\partial_t
\Srad{i}+\partial_j(\alpha\sg{R^j}_i)&=&\alpha \sg \left(\frac{1}{2}
R^{\alpha\beta}g_{\alpha\beta,i}- G_i\right) \ , \ \ \ \ \ 
\end{eqnarray}
where the radiation momentum density variable is defined as
\begin{eqnarray}
\Srad{i}&=& \alpha\sg {R^0}_i\cr
&=&\alpha \sg \left( \frac{4}{3} \Erad u^0 u_i +
\Frad^0 u_i + \Frad_i u^0 \right) \ . \label{Srad_def} 
\end{eqnarray}

\subsection{Evolution of large-scale electromagnetic fields}
The evolution equation for the electromagnetic field in a perfectly 
conducting MHD fluid $\left(F^{\mu \nu}u_{\nu} = 0\right)$ 
can be obtained in conservative form by taking the dual of 
Maxwell's equation $F_{\left[ \mu \nu, \lambda \right]} = 0$. One finds
\begin{equation}
\label{maxwell}
\nabla_{\nu} {}^*F^{\mu \nu} = \frac{1}{\alpha \sqrt{\gamma}} 
\partial_{\nu} \left( \alpha \sqrt{\gamma}\  {}^*F^{\mu \nu} \right) = 0 \ ,
\end{equation}
where $F^{\alpha \beta}$ is the Faraday tensor, and  
${}^* F^{\alpha \beta}=\epsilon^{\alpha \beta \mu \nu} F_{\mu \nu}/2$ 
is its dual.  Using the fact that the magnetic field as measured by 
a normal observer $n^{\alpha}$ is given by $B^i = n_{\mu}{}^*F^{\mu i}$, 
the time component of Eq.~(\ref{maxwell}) gives the 
no-monopole constraint $\partial_j \tilde{B}^j = 0$, 
where $\tilde{B}^j = \sg B^j$. The spatial components of 
Eq.~(\ref{maxwell}) give the magnetic induction equation, 
which can be written as 
\begin{equation}
\label{induction}
\partial_t \tilde{B}^i + \partial_j \left( v^j \tilde{B}^i -  v^i \tilde{B}^j\right) = 0 \ ,
\end{equation}
where $v^i \equiv u^i/u^0$.

\subsection{Evolution of the MHD field}
\label{mhd_evolution_section}

In the MHD limit, $\Tem{\mu}{\nu}$ can be expressed as
\begin{equation}
  \Tem{\mu}{\nu} = b^2 u^{\mu} u^{\nu} + \frac{1}{2} b^2 g^{\mu \nu} 
 - b^{\mu} b^{\nu} \ , 
\end{equation}
where $b^{\mu} = B^{\mu}_{(u)}/\sqrt{4 \pi}$ and where
\begin{equation}
B^{\mu}_{(u)} = u_{\nu} {}^*F^{\nu \mu} 
= -\frac{h^{\mu}{}_{\nu}B^{\nu}}{n_{\nu}u^{\nu}}
\end{equation}
is the magnetic field measured by an observer comoving with the 
fluid. 
The stress-energy tensor associated with the perfect fluid can be 
expressed as
\begin{equation}
\Thydro{\mu}{\nu} = \rho_0 h u^{\mu} u^{\nu} + P  
 g^{\mu \nu}  \ ,
\end{equation}
where $\rho_0$ is the (baryon) rest-mass density, $P$ is matter pressure, $h=1+\epsilon + P/\rho_0$ is the specific 
enthalpy, and $\epsilon$ is the specific internal energy density of the matter.
For brevity, we denote
\begin{equation}
\Tmhd{\alpha}{\beta} = \Thydro{\alpha}{\beta} + \Tem{\alpha}{\beta} \ .
\end{equation}
Thus, we see that the conservation of the total stress-energy tensor
can be written as
\begin{equation}
\label{stress_energy_cons}
\Ttotal{\alpha}{\beta}{}_{;\beta}=\left[\Tmhd{\alpha}{\beta}  +
\Trad{\alpha}{\beta} \right]_{;\beta} = 0
\end{equation}
This can be combined with (\ref{rad_evolution}) to give
\begin{equation}
\Tmhd{\alpha}{\beta}{}_{;\beta} = G^{\alpha} \ .
\label{eq:temp0001}
\end{equation}
Additionally, we have the continuity equation expressing baryon number 
conservation,
\begin{equation}
\label{continuity}
(\rho_0 u^{\nu})_{;\nu}=0 \ .
\end{equation}
Rewriting Eqs.~(\ref{eq:temp0001}) and (\ref{continuity}) in conservative form gives 
(cf. Section IIC in~\cite{mhd_code_paper})
\begin{equation}
\label{mhd_continuity}
\partial_t \rho_* + \partial_j (\rho_* v^j) = 0 \ ,
\end{equation}
\begin{equation}
\label{mhd_momentum}
\partial_t \Smhd{i} + \partial_j (\alpha \sqrt{\gamma} \Tmhdlow{j}{i})
= \frac{1}{2} \alpha \sg \Tmhd{\alpha}{\beta} g_{\alpha \beta,i}+\alpha \sg G_i
\ ,
\end{equation}
\begin{equation}
\label{mhd_energy}
 \partial_t \taumhd + \partial_i ( \alpha^2 \sqrt{\gamma} \Tmhd{0}{i} 
-\rho_* v^i) = s + \alpha^2 \sqrt{\gamma} G^0 \ ,
\end{equation}
where the MHD evolution variables are
\begin{eqnarray}
\rho_* &=& \alpha \sg \rho_0 u^0 \label{eq:rho_star}  \ ,\\
\Smhd{i} &=& \sg n_{\mu} \Tmhdlow{\mu}{i} \cr
&=& \alpha \sg \Tmhdlow{0}{i} \cr 
  &=& (\rho_* h + \alpha u^0 \sg b^2) u_i - \alpha \sg b^0 b_i \ , \\
 \taumhd &=& \sqrt{\gamma}\, n_{\mu} n_{\nu} \Tmhd{\mu}{\nu} - \rho_*\cr
&=& \alpha^2 \sqrt{\gamma}\, \Tmhd{0}{0} - \rho_* \ , \label{eq:taumhd}
\end{eqnarray}
and the source term $s$ is
\begin{eqnarray}
s &=& -\alpha \sqrt{\gamma}\, \Tmhd{\mu}{\nu} \nabla_{\nu} n_{\mu}  \cr
   &=& \alpha \sqrt{\gamma}\, [ (\Tmhd{0}{0}\beta^i \beta^j + 2 \Tmhd{0}{i} \beta^j 
+ \Tmhd{i}{j}) K_{ij} \cr
 & & - (\Tmhd{0}{0} \beta^i + \Tmhd{0}{i}) \partial_i \alpha ] \ .
\end{eqnarray}
Note that these evolution variables are very similar to those 
in~\cite{mhd_code_paper}. The only difference is that there are new radiative 
source terms $G_i$ and $G^0$ in the momentum and energy equations 
(\ref{mhd_momentum}) and (\ref{mhd_energy}), respectively.

To complete the system of equations, it remains only to specify the
equation of state (EOS) of the fluid. 
In this paper, we adopt a $\Gamma$-law EOS,
\begin{equation}
\label{ideal_P}
P = (\Gamma - 1)\rho_0\epsilon,
\end{equation}
where $\Gamma$ is the adiabatic gas constant.
We choose a $\Gamma$-law EOS because it simplifies some of the 
calculations, it is applicable to many cases of interest, and it is 
a standard choice for demonstrating new computational techniques
in the numerical relativity literature. Also, the analytic solutions we are
going to use as code tests also use this
EOS. Nevertheless, all evolution equations derived 
in this section apply for any equation of state, and generalization 
to a more realistic EOS is straightforward. In fact, our code is currently 
capable of handling the general class of EOSs of the form $P=P(\rho_0,\epsilon)$.

The fluid temperature $T$ is required in the radiation force density
term $G^\mu$ (Eq.~(\ref{B_def})). In this paper, we compute it by 
using the ideal gas
law $P=nk_BT=\rho_0k_BT/m$, where $n$ is the baryon number density, 
$m=\rho_0/n$ is the mean mass of 
the baryons in the fluid, and $k_B$ is Boltzmann's
constant. Hereafter we set $k_B=1$.

We point out that the fluid flow is {\em nonadiabatic} in general. In 
particular, there is energy exchange between the matter and radiation 
fields. Also, shocks may be present in some applications.

\subsection{Summary of equations}

To reiterate, the system of coupled Einstein-radiation-Maxwell-MHD equations 
we consider are the BSSN equations (\ref{evolve_gamma})--(\ref{evolve_Gamma}), 
the radiation transport equations (\ref{rad_energy}) and (\ref{rad_momentum}), 
the magnetic induction equation (\ref{induction}), and the MHD equations 
(\ref{mhd_continuity})--(\ref{mhd_energy}). In 
Appendix~\ref{newtonian_appendix}, we demonstrate that our equations 
reduce to the more familiar Newtonian form
in the weak-field, slow-velocity limit. The evolution variables 
are $\phi$, $\tilde{\gamma}_{ij}$, $K$, $\tilde{A}_{ij}$, $\tilde{\Gamma^i}$, 
$\taurad$, $\Srad{i}$, $\tilde{B}^i$, $\rho_*$, $\Smhd{i}$ and $\taumhd$. 
These variables are not completely independent: the BSSN variables 
$\phi$, $\tilde{\gamma}_{ij}$, $K$, $\tilde{A}_{ij}$, and 
$\tilde{\Gamma^i}$ have to satisfy the Hamiltonian constraint 
(\ref{Hamiltonian_BSSN}) and the momentum constraint (\ref{momentum_BSSN}); 
the magnetic field variables $\tilde{B}^i$ have to satisfy the no-monopole 
constraint $\partial_i \tilde{B}^i=0$.

The total stress-energy tensor $T^{\mu \nu}$ is given by 
\begin{eqnarray}
\label{stress_energy_total}
\Ttotal{\mu}{\nu} &=& \Thydro{\mu}{\nu}+\Tem{\mu}{\nu}
+\Trad{\alpha}{\beta} \cr
 && \cr
 &=& \left(\rho_0h + b^2 + \frac{4}{3}\Erad \right) u^{\mu} u^{\nu} 
  + \left( P + \frac{1}{2}b^2 + \frac{1}{3} \Erad \right) g^{\mu \nu} \cr
  && \cr
  && + \Frad^{\mu} u^{\nu} + \Frad^{\nu} u^{\mu} - b^{\mu} b^{\nu} \ . 
\end{eqnarray}
The BSSN matter-energy source terms [Eq.~(\ref{source_def})] can 
be expressed as 
\begin{eqnarray}
  \rho &=& (\alpha u^0)^2 \left(\rho_0h + b^2 + \frac{4}{3}\Erad \right) 
- \left( P + \frac{1}{2}b^2 + \frac{1}{3}\Erad \right) \cr
  && \cr
  && + 2\alpha^2 u^0 \Frad^0 - (\alpha b^0)^2 \\ 
  && \cr
  S_i &=& \alpha u_0 \left( \rho_0h + b^2 + \frac{4}{3}\Erad \right) u_i 
 + \alpha \Frad^0 u_i + \alpha u^0 \Frad_i \cr
  && \cr 
  && -\alpha b^0 b_i \\
  S_{ij} &=& \left(\rho_0h + b^2 + \frac{4}{3}\Erad \right) u_i u_j 
  + \left( P + \frac{1}{2}b^2 + \frac{1}{3} \Erad \right) \gamma_{ij} \cr 
  && \cr
  && + \Frad_i u_j + \Frad_j u_i - b_i b_j \ .
\end{eqnarray}

\section{Implementation}
\label{implementation}

We use a cell-centered Cartesian grid in our three-dimensional simulations. 
Sometimes, symmetries can be invoked to reduce the integration domain. For
octant symmetric systems, we evolve only the upper octant; for
equatorially symmetric systems, we evolve only the upper half-plane.  For
axisymmetric systems, we evolve only the $x$-$z$ plane (a 2+1 
dimensional problem). 
In axisymmetric evolutions, 
we adopt the Cartoon method~\cite{cartoon} for evolving the BSSN equations, 
and use a cylindrical grid for evolving the induction, MHD, 
and radiation equations~\cite{dlss04}.

Our code uses the Cactus parallelization framework
\cite{Cactus}, with the time-stepping algorithm based on 
the {\tt MoL}, or method of lines, thorn. 
In the metric evolution (BSSN sector), spatial derivatives 
can be calculated using second-order or fourth-order finite 
differencing schemes. The Cactus {\tt MoL} thorn allows us to switch to 
a higher-order time-stepping scheme easily. Higher order schemes 
are very useful for evolving spacetimes containing black holes 
using the moving puncture techniques (see, e.g.,~\cite{rit06,goddard06}). 
However, we do not treat puncture black holes here and 
we are currently using a HRSC scheme which is at most 
second-order accurate to evolve the Maxwell, radiation, and MHD equations. 
Hence we use second-order 
finite differencing scheme in the BSSN sector and 
(second-order) iterated Crank-Nicholson 
time-stepping in our calculations.

Our technique for metric evolution is described in our
earlier papers~\cite{dmsb03,dsy04,junk-ID}, so we focus here on
our MHD, induction and radiation algorithms.
The goal of this part of the numerical evolution is to determine the
fundamental ``primitive'' variables ${\bf P}\equiv 
(\rho_0,P,v^i,B^i,\Erad,\Frad^i)$ at future times, given
initial values of ${\bf P}$. 
The evolution equations
(\ref{rad_energy}), (\ref{rad_momentum}), (\ref{induction}),
(\ref{mhd_continuity}), (\ref{mhd_momentum})
and (\ref{mhd_energy}) are written in conservative form:
\begin{equation}
\partial_t {\bf U} + \nabla\cdot{\bf F} = {\bf S}\ ,
\label{cons}
\end{equation}
where the evolution variables 
${\bf U}({\bf P})$ $\equiv$  ($\rho_*$,$\tilde\tau$,$\tilde{S}_i$,$\tilde{B}^i$,$\taurad$,$\Srad{i}$),  
the fluxes ${\bf F}({\bf P})$ and the sources ${\bf S}({\bf P})$ are not explicit
functions of derivatives of the primitive variables, although
they are explicit functions of the metric and its derivatives.
As mentioned above, we evolve Eq.~(\ref{cons})
using the iterated Crank-Nicholson scheme.  
This scheme is second order in time and will be stable if 
$\Delta t < \min(\Delta x^i)/c_{\rm max}$,
where in our case $c_{\rm max}$ is the speed of light. For each Crank-Nicholson
substep, we first update the gravitational field variables (the BSSN 
variables).  We then update the electromagnetic fields $B^i$ by integrating 
the induction equation. 
Next, the  MHD variables ($\rho_{\star}$, $\tilde{\tau}$, and
$\tilde{S}_i$) are updated.  Then we update the radiation variables
($\taurad$, and $\Srad{i}$). Finally, we use these updated values to recover
the primitive variables on the new timestep. Below, we briefly summarize 
some of the important techniques we utilize during the evolution.

\subsection{Reconstruction step}
\label{sec:reconstruction}
We implement an
approximate Riemann solver to handle the advection in Eq.~(\ref{cons}). 
For simplicity, we consider the one-dimensional case here. The generalization 
to multi-dimension is straightforward.
The first step in calculating this flux is to compute ${\bf P}_L =
{\bf P}_{i+1/2-\epsilon}$ and 
${\bf P}_R = {\bf P}_{i+1/2+\epsilon}$, i.e.\ the primitive variables
to the left and right of the grid cell interface. 
As in~\cite{mhd_code_paper}, we use the Monotonized central (MC) 
scheme~\cite{mc} to compute the primitive variables at the cell interface.  
This scheme is second-order accurate at most points when the data are smooth, 
but becomes first-order accurate across a discontinuity (e.g.\ shock).
(See~\cite{mhd_code_paper} for other reconstruction methods in
our MHD code.)

\subsection{Riemann solver step}
\label{sec:riemann}
Next, we take the reconstructed data as initial data for a
piecewise constant Riemann problem, with ${\bf P} = {\bf P}_L$ on
the left of the interface, and ${\bf P} = {\bf P}_R$ on
the right of the interface.  The net flux at the cell interface
is given by the solution to this Riemann problem.

We use the HLL (Harten, Lax, and van Leer) approximate Riemann
solver~\cite{HLL}. Our implementation has been described in
\cite{mhd_code_paper}.  To summarize, HLL fluxes are given by
\begin{equation}
\label{eq:hll}
f_{i+1/2} = \frac{c_{\rm min} f_R + c_{\rm max} f_L
  - c_{\rm min}c_{\rm max} (u_R - u_L)}{c_{\rm max} + c_{\rm min} }\ .
\end{equation}
Here
\begin{eqnarray}
c_{\rm max} &\equiv& \max(0,c_{+R},c_{+L})\cr
c_{\rm min} &\equiv& -\min(0,c_{-R},c_{-L})
\end{eqnarray}
where $c_+$ is the maximum right-going wave speed and $c_-$ is the
maximum left-going wave speed. 
We obtain $c_{\pm}$ by solving the dispersion relation for waves with
wave vectors of the form
\begin{equation}
k_{\mu} = (-\omega,k_1,0,0)
\end{equation}
The wave speed is simply the phase speed $\omega/k_1$. We find the
dispersion relation in the comoving frame of
the fluid (denoted by subscript cm), and hence
$\omega_{\rm cm}/k_{\rm cm}$,
as described in Appendix~\ref{char_appendix}.
To obtain $\omega/k_1$ in the grid frame, we use the dispersion 
relation~(\ref{eq:dispersion}) and substitute the $\omega_{\mathrm{cm}}
= - k_{\mu} u^{\mu}$, and $k_{\mathrm{cm}}^2 = K_{\mu}K^{\mu}$, where
$K_{\mu} = (g_{\mu \nu} + u_{\mu} u_{\nu})k^{\nu}$. Wave speeds in the
$y$-direction and $z$-direction are found analogously.\\

\subsection{Recovery of primitive variables}
\label{sec:primitives}
Having computed ${\bf U}$ at the new timestep, we must use these values to
recover ${\bf P}$, the primitive variables on the new time level. We
can recover the hydrodynamics primitive variables $\rho_0,P,v^i$
from the MHD evolution variables $\rhostar,\taumhd,\Smhd{i}$
numerically, as described in Section~III~C
in~\cite{mhd_code_paper}. Once the fluid velocity $v^i$ is found, the 
radiation primitive variables $\Erad$ and $\Frad^i$ can be computed from the 
radiation evolution variables $\taurad$ and $\Srad{i}$ analytically 
using Eqs.~(\ref{taurad_def}) and~(\ref{Srad_def}). We solve the
following set of two coupled linear equations to recover $\Erad$ and $\Frad^0$:
\begin{equation}
\taurad=\sqrt{\gamma}\left[ \left( \frac{4}{3}(\alpha
u^0)^2 - \frac{1}{3} \right) \Erad +
2\alpha^2 u^0 \Frad^0 \right]
\end{equation}
\begin{equation}
-\alpha
 u^0\taurad+(u^0\beta^i+u^i)\Srad{i}=-\alpha\sqrt{\gamma}(E u^0+\Frad^0)
\end{equation}
The first one is just Eq.~(\ref{taurad_def}), while the
second one is obtained by using 
$u_\mu F^\mu=0$ to eliminate $F_i$ in
Eq.~(\ref{Srad_def}). After solving for $\Erad$ and $\Frad^0$, we compute 
$\Frad^i$ by
\begin{equation}
\Frad^i=\frac{\gamma^{ij}\Srad{j}}{\alpha\sqrt{\gamma}u^0} -
\frac{4}{3} E u^0\left(v^i+\beta^i\right)-2\Frad^0\beta^i-\Frad^0v^i \ ,
\end{equation}
which is derived by raising the index of $F_i$ in Eq.~(\ref{Srad_def}).

\subsection{Constrained Transport}
The Maxwell equation demands that the magnetic fields $\tilde{B}^i$ 
satisfy the no-monopole constraint $\partial_i \tilde{B}^i = 0$. 
Unphysical behavior may arise if this constraint is violated.
Thus, ``constrained transport schemes'' have been
designed to evolve the induction equation while maintaining 
$\partial_i \tilde B^i = 0$ to roundoff precision~\cite{eh88}.  We use the
flux-interpolated constrained transport (flux-CT) scheme introduced by
T\'oth~\cite{t00}
and used by Gammie {\it et al}~\cite{HARM}.  This scheme involves
replacing the induction equation flux computed at each point with
linear combinations of the fluxes computed at that point and neighboring
points.  The combination assures both that second-order accuracy is
maintained, and a particular finite-difference representation of 
$\partial_i\tilde B^i = 0$ is enforced to machine precision.

\subsection{Low-density regions and boundary conditions}

\subsubsection{Low-density regions}

If vacuum exists anywhere in our computational domain, the MHD
approximation will not apply in this region, and we will have to solve
the vacuum Maxwell equations there (see e.g.~\cite{bs03b}).  In 
addition, the optically thick assumption on the radiation field 
in Sec.~\ref{sec:radiation-fields} also breaks down in sufficiently 
low density regions. 
In many astrophysical scenarios, however, a sufficiently dense, 
ionized plasma will exist outside the stars or disks, whereby MHD 
will remain valid in its force-free limit.  A similar situation may
arise for the radiation field, where the ambient gas in our
computational domain may be sufficiently dense to maintain an optical
depth above unity. However, in some applications we may need to take
into account the transition from optically thick to optically thin limits in 
the low-density regions, depending on the magnitude of the
opacity. A precise treatment of this problem requires solving the full
Boltzmann radiative transfer equation
(see e.g.~\cite{stu_rad_2,rj02,lmtmhb01,bypet00}; See also
\cite{lp81,lwf07} for approximation schemes.) In this
paper, however, we avoid this issue. 
For the code tests that do not have low density regions 
(Sec.~\ref{sec:1d-test}),  no special treatment is required.
We do, however, present a test involving Oppenheimer-Snyder collapse 
(Sec.~\ref{sec:os}) where there is a vacuum outside the star. 
As in many hydrodynamics 
simulations in astrophysics, we impose a low-density ``atmosphere'' 
outside the star to facilitate the integration of hydrodynamics
equations. It turns out that our atmosphere scheme suffices to mimic
the correct (``zero temperature'') radiation boundary conditions that
we wish to impose at
the surface of the star (see Sec.~\ref{sec:os}).

In the low-density regions near the 
surface of the star, we sometimes encounter problems when 
recovering the primitive variables; in particular, the 
equations ${\bf U} = {\bf U}({\bf P})$ occasionally have no physical solution. 
Usually, unphysical ${\bf U}$ are those values corresponding to negative pressure. 
As in~\cite{mhd_code_paper}, we apply a fix at these points, first 
suggested by Font {\it et al}~\cite{fmst00}. In the system of equations 
(\ref{eq:rho_star})--(\ref{eq:taumhd}) 
to be solved, we replace Eq.~(\ref{eq:taumhd})
with the adiabatic relation $P = \kappa\rho_0^{\Gamma}$, where $\kappa$ is set 
equal to its initial value.  This substitution guarantees a positive
pressure. Typically, these low density regions have little influence
on the dynamical evolution of the system, which is the principal
target of our current investigations.

\subsubsection{Boundary conditions}

For the code test in Sec.~\ref{sec:1d-test}, we evolve one-dimensional,  
radiation-hydrodynamics equations in a fixed, Minkowski spacetime. We impose 
the ``copy'' boundary condition on all the evolution variables, i.e.\ 
variables at the boundaries are copied from the closest grid point.

For the Oppenheimer-Snyder code test in Sec.~\ref{sec:os}, we evolve the system 
of coupled Einstein-radiation-hydrodynamics equations. In this case, 
we employ Sommerfeld outgoing wave boundary 
conditions for all BSSN and gauge variables:
\begin{equation}
\label{wavelike}  
f(r,t) = {r - \Delta r\over r}f(r-\Delta r, t-\Delta T),
\end{equation}
where $\Delta T$ is the timestep and $\Delta r = \alpha
e^{-2\phi}\Delta T$. 
For the radiation hydrodynamics, we impose the outflow boundary condition 
on the primitive variables $\rho$, $P$, $v^i$, $\Erad$, and $\Frad^i$ (i.e., the variables are 
copied along the grid directions with the condition that the velocities 
be positive or zero in the outer grid zones).  We note that 
the radiation in this test is initially confined inside the 
star, but escapes from the stellar surface during the evolution. In
the end of the simulation, the total emitted radiation remains
small and the dynamics  
of the system is insensitive to the boundary condition employed.

\section{Code Tests}
\label{tests}
Our GRMHD code has previously been thoroughly tested by maintaining
stable rotating stars in stationary equilibrium, by reproducing
Oppenheimer-Snyder collapse to a black hole, and by reproducing
analytic solutions involving MHD shocks, nonlinear
MHD wave propagation, magnetized Bondi accretion, and MHD waves induced
by linear gravitational waves \cite{mhd_code_paper}. It has also been
compared with the GRMHD code of Shibata and Sekiguchi \cite{ss05} by
performing simulations of the evolution of magnetized hypermassive
neutron stars \cite{dlsss06a,dlsss06b}, and of magnetorotational collapse of stellar
cores \cite{slss06}.  We obtain good agreement between these two independent
codes.   Our code has also been used to study the evolution of BHBH
and BHNS binaries~\cite{eflstb07}, and the
evolution of relativistic hydrodynamic matter in the presence of 
puncture black holes~\cite{fbest07}. Here we restrict our attention to
testing the new
radiation-hydrodynamics sector, setting large-scale magnetic fields to
zero. We also choose the grey body absorption opacity $\kappa^a$ to be
a constant and set the scattering opacity $\kappa^s$ to zero.
\subsection{Minkowski radiation-hydrodynamics tests}
\label{table1}
\begin{table*}
\caption{Initial states for one-dimensional tests.}
\begin{tabular}{l l l l l l l}
\hline \hline
Test & $\Gamma$ & $\kappa^a$ & Left state${}^{\rm c}$ & Right State${}^{\rm c}$ & $t_{\rm
  final}$ \ \  & $t_{\rm{sc}}{}^{\rm b}$ \\ 
\hline \hline
1 & \ \ 5/3 \ \ & 0.4 \ \  & $\rho_0=1.0$ \ \ & $\rho_0=2.4$ \ \ & 5000&2000\\
($\mu=0.0){}^{\rm a}$ &&& $P=3.0 \times 10^{-5}$  & $P=1.61 \times 10^{-4}$  &&\\
& && $u^x=0.015$ & $u^x=6.25 \times 10^{-3}$ & & \\
&&& $\Erad=1.0 \times 10^{-8}$ & $\Erad=2.51 \times 10^{-7}$ &  &\\
\hline
2 & 5/3 & 0.2 & $\rho_0=1.0$  & $\rho_0=3.11$  & 100& 80\\
($\mu=0.1){}^{\rm a}$ &&& $P=4.0\times 10^{-3}$  & $P=0.04512$  & &\\
&&& $u^x=0.25$ & $u^x=0.0804$ &  \\
&&& $\Erad=2.0\times 10^{-5}$ & $\Erad=3.46 \times 10^{-3}$ &  &\\
\hline
3 & 2 & 0.3 & $\rho_0=1.0$  & $\rho_0=8.0$  & 20&20\\
($\mu=0.8){}^{\rm a}$ &&& $P=60.0$  & $P=2.34\times 10^{3}$  &&\\
&&& $u^x=10.0$ & $u^x=1.25$ &  \\
&&& $\Erad=2.0$ & $\Erad=1.14\times 10^{3}$ & & \\
\hline
4 & 5/3 & 0.08 & $\rho_0=1.0$  & $\rho_0=3.65$  & 100&90\\
($\mu=0.1){}^{\rm a}$ &&& $P=6.0\times 10^{-3}$  & $P=3.59\times 10^{-2}$  & &\\
&&& $u^x=0.69$ & $u^x=0.189$ & &\\
&&& $\Erad=0.18$ & $\Erad=1.30$ & &\\
\hline \hline
\end{tabular}
\vskip 12pt
\begin{minipage}{12cm}
\raggedright
${}^{\rm a}$ {$\mu$ is the speed at which the wave travels. Traveling wave solutions are 
obtained by boosting the stationary solutions in
Appendix~\ref{analytic_shock_appendix}} to speed $\mu$. \\
${}^{\rm b}$ {$t_{\rm{sc}}$ is the approximate time it takes for a wave traveling at the
  sound speed to propagate from the center of the grid to the right
  boundary}. \\
${}^{\rm c}$ {Values refer to asymptotic regions.  We
  solve ODEs to determine the exact solution in the transition region (see Appendix \ref{analytic_shock_appendix}).} 
\end{minipage}
\label{tab:1Dtests}
\end{table*}

\label{sec:1d-test}
We present here a series of tests of nonlinear
  radiation-hydrodynamic waves in Minkowski spacetime with planar symmetry.  These tests
  are summarized in Table~\ref{tab:1Dtests}.  For our initial data, we 
  generate semi-analytic, stationary configurations using the method outlined in
  Appendix~\ref{analytic_shock_appendix}. To test the ability of our code 
  to handle shocks and waves moving across the grid, we boost the stationary 
  solutions derived in Appendix~\ref{analytic_shock_appendix} for tests 2--4.  
  In each case, our
  computational domain is $x \in (-20,20)$. We choose the opacities in
  each case to ensure that the grid boundaries at $x=\pm 20$ reside in the asymptotic region where 
  all hydrodynamic and radiation quantities approach their asymptotic 
  values, and that the total optical depth across the grid is $\tau \sim 10$ (see Appendix~\ref{analytic_shock_appendix}). 

  We evolve the system with a timestep $\Delta t = \Delta x$ for test 1 and test 2, and $\Delta t =
  0.1 \Delta x$ for test 3 and test 4.  
  We use resolutions ranging from $\Delta x = 0.0125$ to $\Delta x = 0.1$ 
  in order to perform convergence tests. To demonstrate convergence,  we consider
a grid function $g$ with error $\delta g = g - g^{\rm exact}$.
We calculate the L1 norm of $\delta g$ (the ``average'' of                                                                                     
$\delta g$) by summing over every grid point $i$:	 
\begin{equation} 
  L1(\delta g) \equiv \Delta x \sum_{i=1}^N |g_i-g^{\rm exact} (x_i)|  \ ,                                                                                                                                   
\end{equation}   
where $N \propto 1/\Delta x$ is the number of grid points.
We find that for our continuous
  configurations (tests 3 and 4), we achieve second order convergence.
  Because our shock capturing scheme becomes 1st order when
  discontinuities are present, we achieve the expected first order
  convergence for our discontinuous configurations (tests 1 and 2). 

  The initial configurations listed in Table~\ref{tab:1Dtests} are chosen to test our code in a variety of
  regimes, including gas-pressure dominated, radiation-pressure
  dominated, Newtonian, relativistic, continuous, and discontinuous
  matter and radiation profiles.  In each test, the equation of state of the gas is given by a $\Gamma$-law EOS.  We choose $\Gamma=5/3$ for each test except our highly-relativistic case, in which we choose $\Gamma=2$. 
This latter choice is adopted because the sound speed ($c_s = \sqrt{\Gamma P/\rho_0h}$) 
for a $\Gamma$-law EOS is limited by $c_s<\sqrt{\Gamma-1}$. 
Highly-relativistic sound speeds 
($c_s \rightarrow 1$) can only be achieved for $\Gamma \geq 2$.
Below, we provide a brief description of each test.

\begin{figure}
\includegraphics[width=8cm]{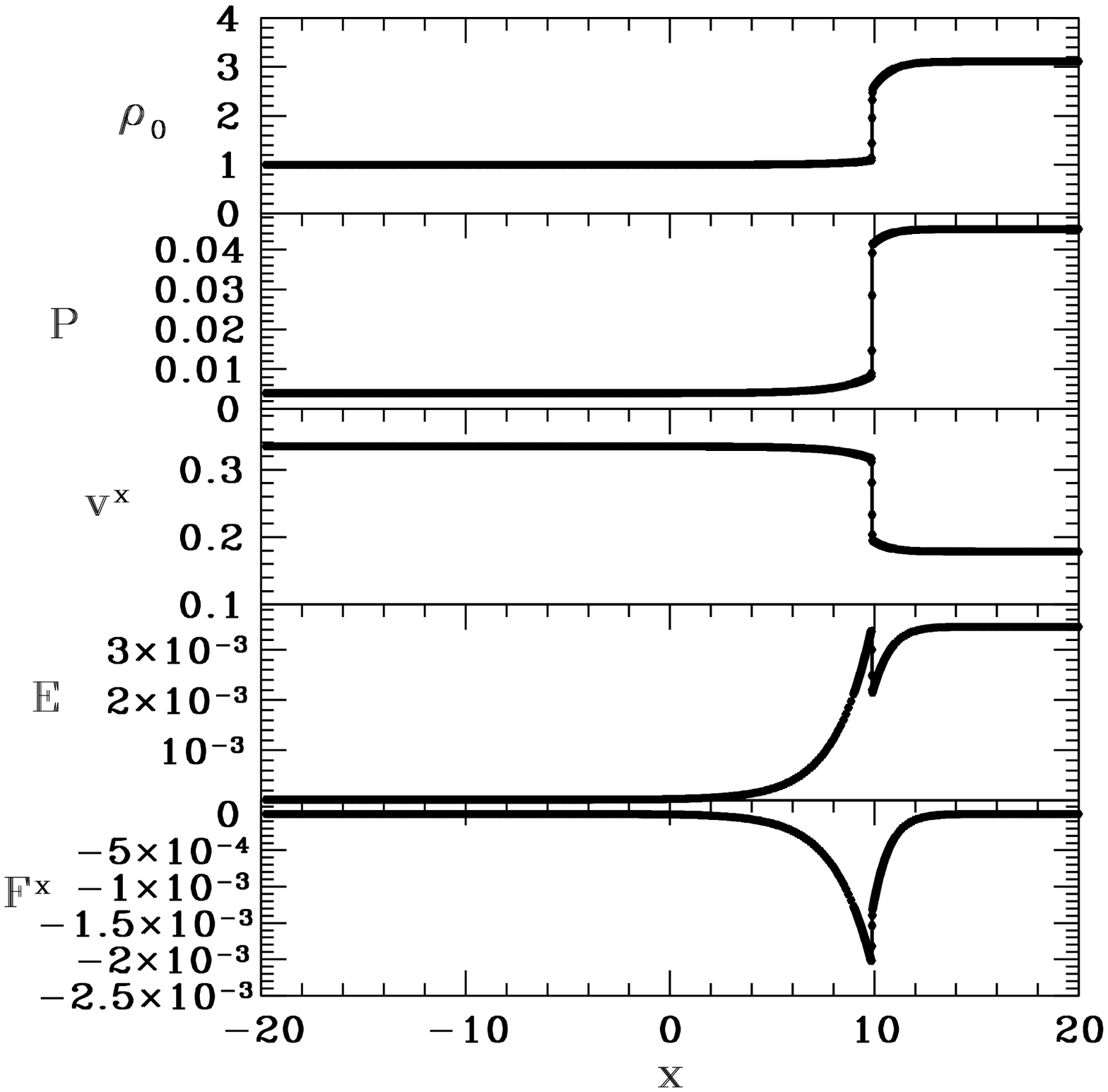}
\caption{Profiles of $\rho_0$, $P$, $v^x$, $\Erad$, and $\Frad^x$ at
$t=100$ for test~2. In this test, the shock front moves with velocity
$\mu=0.1$.  Solid dots denote data from numerical simulations with resolution $\Delta x = 0.0125$.  Solid lines denote the exact solutions (Appendix~\ref{analytic_shock_appendix}).}
\label{fig:discont2_profile}
\end{figure}

\begin{figure}
\includegraphics[width=8cm]{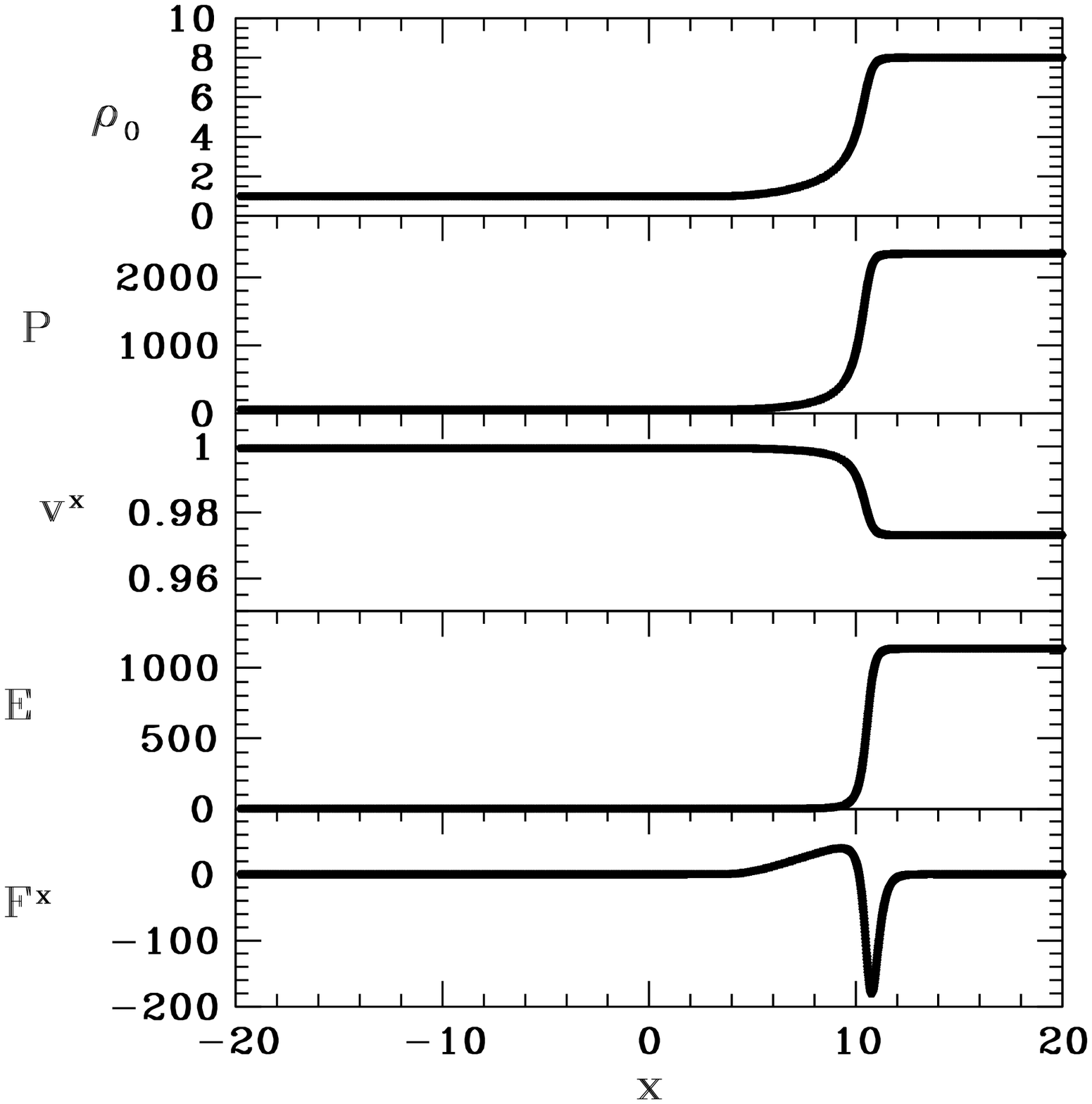}
\caption{Profiles of $\rho_0$, $P$, $v^x$, $\Erad$, and $\Frad^x$ at
$t=20$ for test~3. In this test, the shock front moves with velocity
$\mu=0.8$.  Solid dots denote data from numerical simulations with resolution $\Delta x = 0.0125$.  Solid lines denote the exact solutions (Appendix~\ref{analytic_shock_appendix}).}
\label{fig:cont3_profile}
\end{figure}

\begin{figure} 
\includegraphics[width=8cm]{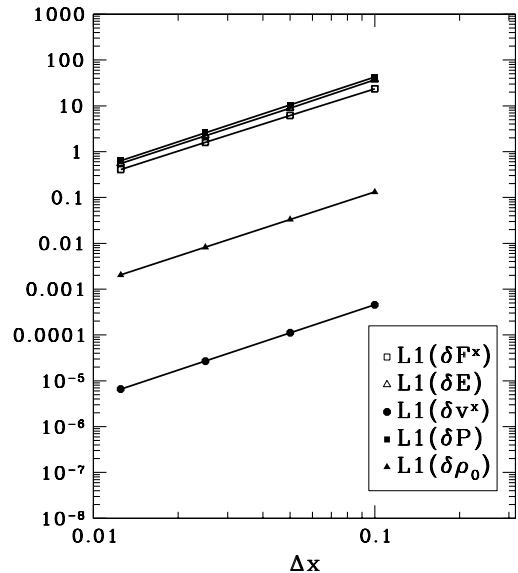}
\caption{L1 norms of the errors in $\rho_0$, $P$, $v^x$, $\Erad$, and
$\Frad^x$ for test 3 at $t=20$.  This log-log plot
shows that the L1 norms of the errors in all quantities are
proportional to $(\Delta x)^2$, and are thus second-order convergent.}  
\label{fig:cont3_conv}
\end{figure}

\begin{figure}
\includegraphics[width=8cm]{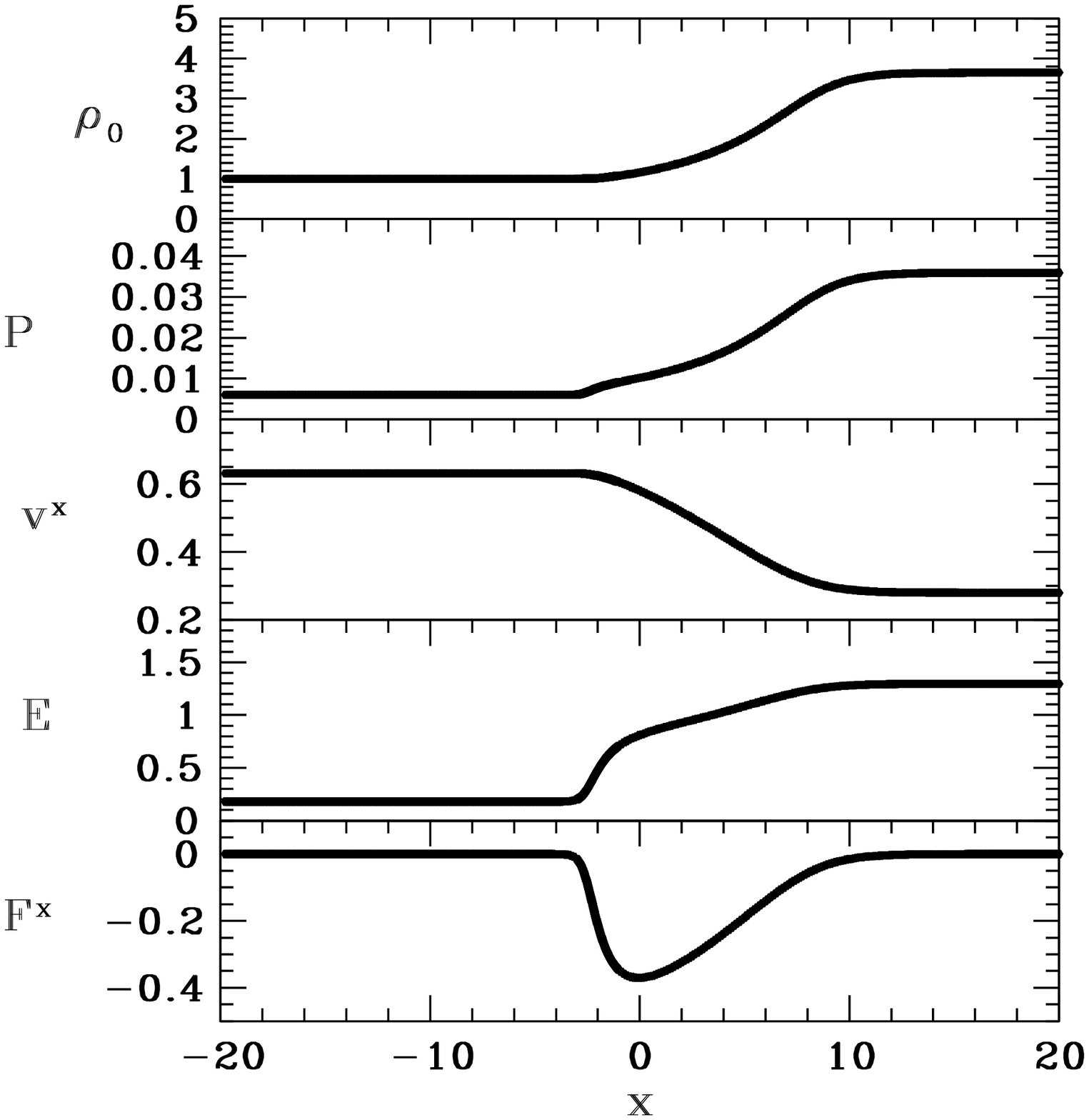}
\caption{Profiles of $\rho_0$, $P$, $v^x$, $\Erad$, and $\Frad^x$ at $t=100$ for test~4.  In this test, the shock front moves with velocity
$\mu=0.1$.  Solid dots denote data from numerical simulations with resolution $\Delta x = 0.0125$. Solid lines denote the exact solutions (Appendix~\ref{analytic_shock_appendix}).}
\label{fig:cont4_profile}
\end{figure}

\begin{enumerate}
\item {\it Nonrelativistic strong shock.}
For this test, we set up a strong, gas-pressure dominated, Newtonian
($u^x_{\mathrm{max}} = 0.015 \ll 1$) shock propagating into a
cold gas.  We have chosen to simulate this scenario because it can be
compared to the analytic solution for a subcritical radiating
shock first derived by Zel'dovich and Raizer \cite{zeldovich_raizer}
and summarized in \cite{mihalas}.  We find very good agreement with
this analytic result (see Fig.~\ref{fig:discont1_profile}).
We note that the radiative shock 
junction conditions (see Appendix~\ref{analytic_shock_appendix}) require that $R^{0x}$ and $R^{00}$ be continuous 
at the shock front, even though $\Erad$ and $\Frad^x$ are, in general, discontinuous 
at the shock.  In the Newtonian limit, however, the continuity of $R^{0x}$ 
and $R^{00}$ is equivalent to the continuity of $\Erad$ and $\Frad^x$.
\item {\it Mildly-relativistic strong shock.}
In this test, we set up a mildly-relativistic ($u^x_{\mathrm{max}}=0.25$), gas-pressure dominated shock.
In this case, we see that $\Erad$ and $\Frad^x$ no longer appear
continuous.  We boost this shock so that the shock speed is $\mu = 0.1$.  We
find that the discontinuity is able to retain its shape very well as
the shock travels and matches very well with the analytic solution
(see Fig.~\ref{fig:discont2_profile}).
\item {\it Highly-relativistic wave.}
In this test, we simulate a highly-relativistic ($u^x_{\mathrm{max}}=
10$), gas-pressure dominated configuration in which
all quantities are continuous, but asymptote to different values on
either side of the computational domain.  We boost this configuration
so that it travels across the grid with velocity $\mu = 0.8$. 
The numerical results agree very well with semi-analytic solution 
(see Fig.~\ref{fig:cont3_profile}).
Figure~\ref{fig:cont3_conv} shows the L1 norms of the errors 
in $\Frad^x$, $\Erad$, $v^x$, $P$ and $\rho_0$ at $t=t_{\rm final}=20$. 
We find that all errors converge to zero at second order in $\Delta x$.
\item {\it Radiation-pressure dominated, mildly-relativistic wave.}
In this test, we study the performance of our code in the radiation-pressure dominated ($P \ll \Prad$), mildly-relativistic
($u^x_{\mathrm{max}} = 0.69$) regime. We boost this configuration
so that it travels across the grid with velocity $\mu = 0.1$. The numerical results again agree with the semi-analytic solution (see Fig.~\ref{fig:cont4_profile}).
\end{enumerate}


\begin{figure}
\includegraphics[width=8cm]{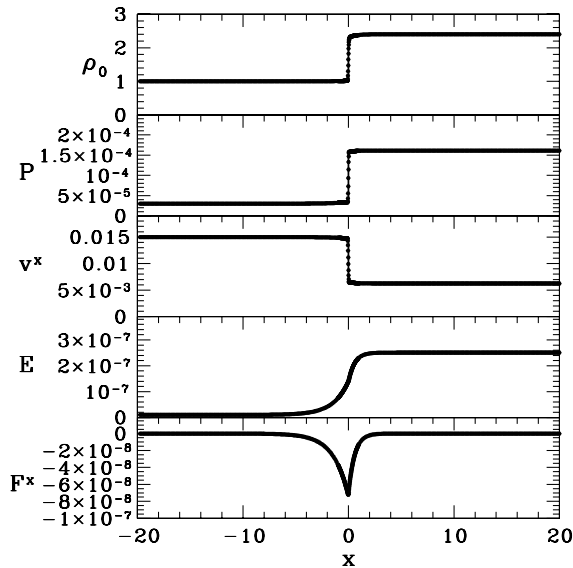}
\caption{Profiles of $\rho_0$, $P$, $v^x$, $\Erad$, and $\Frad^x$ at
$t=5000$ for test 1. In this test, the shock front remains
stationary.  Solid dots denote data from numerical simulations with resolution $\Delta x = 0.0125$.  Solid lines denote the exact solutions (Appendix~\ref{analytic_shock_appendix}).}
\label{fig:discont1_profile}
\end{figure}

\subsection{Dynamical Spacetime test: Thermal Oppenheimer-Snyder Collapse}
\label{sec:os}
The collapse from rest of a
homogeneous dust ball ($P=0$) in general relativity can be described
by the analytic Oppenheimer-Snyder solution \cite{OS39}. The collapse
results in the formation of a Schwarzschild black hole.
The evolution of thermal
radiation within the dust ball has been considered by Shapiro
in~\cite{stu_rad_1,stu_rad_2}. In both papers, the radiation is
assumed to be a small perturbation, so that the dynamics are
unaffected by the presence 
of radiation, and the matter and metric profiles can still be
described by the Oppenheimer-Snyder solution. The first paper employs
the relativistic thermal diffusion approximation for the
radiation, and derives analytic solutions for both the Newtonian and
general relativistic 
cases. In the second paper this approximation is removed and
replaced by solving the exact radiative transfer (Boltzmann)
equation for the intensity, coupled to the radiation moment equations
for the radiation flux and energy density. It is found that 
the results obtained by solving the Boltzmann transport equation agree very 
well with the analytic solutions assuming diffusion approximations,
provided the optical depth of the star is sufficiently large ($\gg$ 1). 
Here we perform a numerical simulation of ``thermal Oppenheimer-Snyder
collapse'' using our radiation GRMHD code, and compare
our results to the analytic solution in the diffusion approximation
limit given in~\cite{stu_rad_1,stu_rad_2}. For convenience, we summarize the 
analytic solution in Appendix~\ref{thermal_os_appendix}.

For our initial data, the areal radius of the star is set to be $R_i=3M$, 
where $M$ is the ADM mass of the star. We choose the
initial profiles for all hydrodynamic and radiation quantities to be
homogeneous throughout the star, in compliance with the analytic
solution in Appendix~\ref{thermal_os_appendix}. 
The analytic solution assumes that (1) the matter and radiation 
pressure is small enough to be dynamically unimportant (i.e.\ 
$P/\rho_0 \ll M/R$ and ${\mathcal P}/\rho_0 =E/3\rho_0\ll M/R$), 
(2) radiation pressure dominates over gas pressure (${\mathcal P}\gg P$), 
(3) gas and radiation are in local thermal equilibrium (LTE), 
i.e.\ $E=4\pi B = a_R T^4$, and (4) the star is optically thick.
To satisfy these conditions, we choose the following initial data: 
$\rho_0=M/(\frac{4}{3}\pi R_i^3)$, $P=10^{-4}\rho_0$, $E=10^{-3}\rho_0$, 
$v^i=0$, $F^i=0$. LTE is achieved in the initial data by fixing the constant
$a_Rm^4=m^4 E/T^4=E(\rho_0/P)^4=10^{13}M/(\frac{4}{3}\pi R_i^3)$. We
note that in our formalism the system is allowed to deviate from 
the LTE during the evolution. However, we find that the system remains 
close to the LTE during the entire evolution and the numerical data agree well 
with the analytic solution (see below). We choose
$\kappa^a$ so that $\tau^a=\kappa^a\rho R=50$ initially. This
guarantees that the star is optically thick initially. As the collapse
proceeds, the optical depth increases as $\tau\propto 1/R^2$, so the star
remains optically thick.

We construct the initial data for the spatial metric by transforming the 
analytic Oppenheimer-Snyder metric from Friedmann to isotropic 
coordinates, following the procedure described in~\cite{OS_Coll}. 
We use the analytic solution only at t=0. The metric at later times 
is evolved, together with hydrodynamics and radiation. 
The lapse and shift are determined by the hyperbolic
driver conditions (Eqs.~(\ref{hb_lapse_nok3}) and
(\ref{hb_shift})). These are gauge conditions that have been widely
used in stellar collapse calculations using the BSSN scheme. We choose
$a_1=0.75$, $b_1=0.15$, $a_2 = b_2 = 
2M^{-1}$, $a_3 = 1$. A smaller $b_1$ prevents ``blowing out'' of the
coordinate system, a well-known 
effect~\cite{sbs00,dmsb03} which can spoil grid resolution in the center of the
collapsing object. We 
perform our numerical simulation in axisymmetry, with $200^2$, 
$400^2$, $800^2$ and $1600^2$ grid points. We choose $\Delta t = 0.1 \Delta x$
in these simulations. The outer boundary is placed at $4M$ in
isotropic coordinates ($R_{out}= 5.06M$ in areal radius). Note that we do not
impose any special boundary 
condition at the stellar surface, in contrast to the zero
temperature boundary condition ($E=0$) used in
the derivation of the analytic solution~\cite{stu_rad_1}. The low
density region outside the star mimics this surface boundary condition, as the
atmosphere is made to be much colder than the interior of the star,
and hence the thermal emission and build-up of radiation energy
density in the atmosphere is negligible. 

The analytic solution given in Appendix~\ref{thermal_os_appendix}
is expressed in Friedmann coordinates (i.e.\ Gaussian normal coordinates comoving 
with the fluid), which is equivalent to using the gauge 
conditions $\alpha=1$ and $\beta^i=0$ (geodesic slicing and zero shift), 
which are different from the gauge conditions we adopt in our numerical 
simulations. In order to compare our
numerical result to the analytic solution, we perform a
mapping between these two different gauges. This is achieved 
first by following a set of Lagrangian fluid elements inside the star, and calculating 
the proper time and position of these elements by integrating the equations
\begin{equation}
\frac{d\tau}{dt}=\frac{1}{u^0},\ \ \ \frac{dx^i}{dt}=v^i .
\end{equation} 
Next, we use the metric and the positions of the fluid elements to
compute their areal radii $r_s$. Finally, knowing the proper times $\tau$ and
areal radii $r_s$ of the fluid elements, we use
Eqs.~(\ref{eq:a_eta}) and (\ref{eq:tau_eta}) and $r_s=a(\tau)\sin\chi$
to compute their Friedmann coordinates ($\tau,\chi$).  The
mapping between these two gauges is thus established. The pair
$(\tau_j,\chi_j)$ for each element $j$ uniquely specifies the fluid
and radiation parameters.

\begin{figure}
\includegraphics[width=8cm]{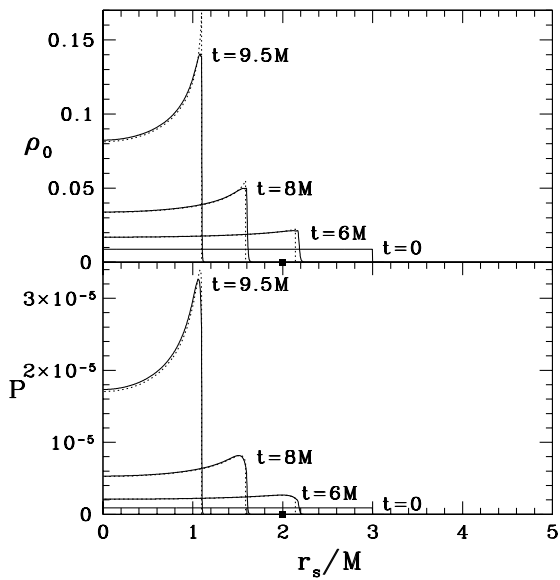}
\caption{Profiles of the hydrodynamic quantities
  $\rho_0$ and $P$ in Schwarzschild (areal) radius at 
  times $t/M$=0, 6, 8 and 9.5 for thermal Oppenheimer-Snyder collapse.
  Solid lines represent numerical data with $1600^2$ grid points, 
  and dashed lines show analytic solutions.
  The solid square on the $x$-axis denotes the radius of the
  apparent horizon, which forms after the stellar surface passes through 
  an areal radius of $2M$.}
\label{fig:hydro_profile}
\end{figure}

\begin{figure}
\includegraphics[width=8cm]{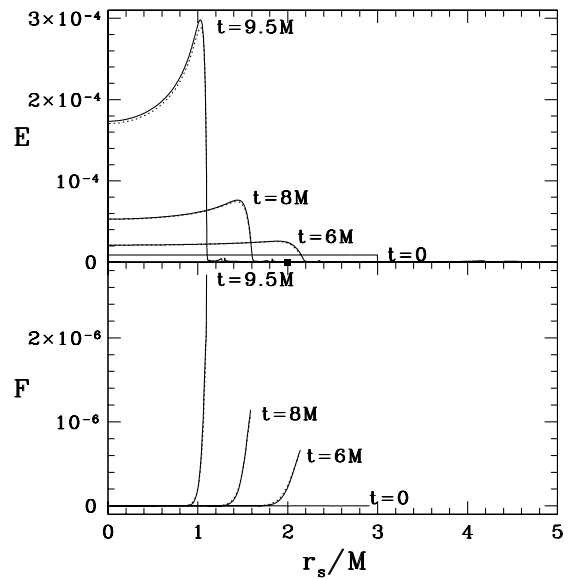}
\caption{Same as Fig.~\ref{fig:hydro_profile} but for the radiation
  quantities $E$ and $F$ inside the star. Note that $F=0$ everywhere at $t=0$.}
\label{fig:rad_profile}
\end{figure}

Figures~\ref{fig:hydro_profile} and~\ref{fig:rad_profile} show the
profiles of $\rho_0$, $P$, $E$ and $F$ at different 
times during the collapse. Note that while the
density remains spatially constant during the collapse in
comoving Friedmann coordinates, this is not true in our gauge.  We see that the numerical results agree very well with the 
analytical solution, even after the apparent horizon appears at $t=6.78M$ and 
all of the stellar material is inside the horizon. 

We next perform a convergence test for the radiation
quantities.  We find that our numerical data converge 
to a solution slightly different from the analytic solution. 
This can be explained by the fact that the analytic solution 
is strictly valid only in the perturbative limits that $P/\rho_0 \rightarrow 0$ and
$P/\Prad \rightarrow 0$. While we set up our initial data to approximate these limits, the slight deviation from the analytic solution is 
still detectable with our resolutions. In the absence of radiation ($E=0$, 
$F^i=0$), we have checked that the deviation in $\rho_0$ between numerical 
and analytic values is reduced by a factor of 10 if we reduce the 
ratio $P/\rho_0$ by a factor of 10. In the presence of radiation, however, 
decreasing the ratios $P/\Prad$ and $\Prad/\rho_0$ 
arbitrarily small makes the numerical simulations quite challenging, as 
accurate evolution for the radiation quantities $E$ and $F^i$ requires 
accurate evolution of the temperature $T\propto P/\rho_0$, which in 
turn requires accurate determination of the pressure $P$. However, accurate 
computation of 
$P$ in the limit $P/\rho_0 \rightarrow 0$ is difficult. Since the evolution variables are dominated by the rest mass density $\rho_0$, in order to recover the
tiny $P$ accurately from them, the numerical
evolution has to be very accurate. This requires very high
resolution. Thus, we perform a convergence test in which we compare
numerical solutions with small but finite $P/\Prad$ and $P/\rho_0$ for different resolutions, rather than
comparing with the analytic solution.

\begin{figure}
\includegraphics[width=8cm]{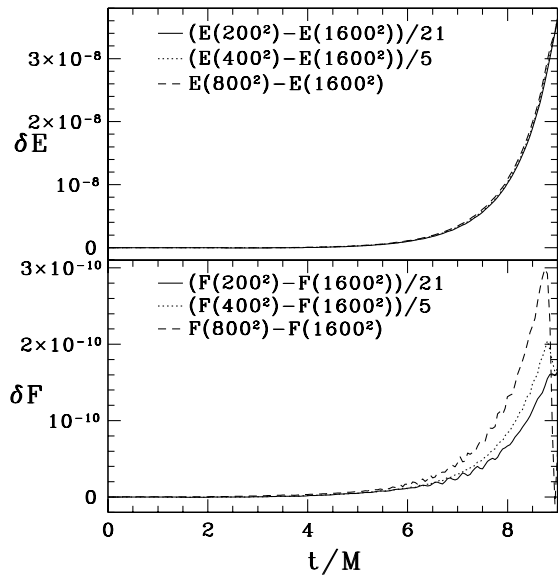}
\caption{Convergence test for the radiation quantities $E$ and $F$, 
  computed at the Lagrangian point halfway
  between the center and the surface of the star (in terms of areal
  radius). Resolutions with $200^2$, $400^2$, $800^2$ and
  $1600^2$ grids are used here. The differences between lower and 
  the highest solution ($1600^2$) are rescaled to demonstrate 
  second-order convergence.} 
\label{fig:os_conv}
\end{figure}

Figure~\ref{fig:os_conv} shows the result of the convergence test for
$E$ and $F$, with the differences scaled for second order
convergence. We follow a Lagrangian fluid element halfway between the
center and the surface of the star (in terms of areal radius),
determine the radiation parameters at the position of the element
versus time, 
and subtract the numerical results from different resolutions. Since 
we use a HRSC scheme that is second-order accurate except at positions 
where discontinuities appear, e.g.\ at the surface of the star where
the density falls abruptly, we expect 
that the order of convergence depends on how much the physical
quantity is affected by the discontinuity at the stellar
surface. In principle, the first-order error will propagate 
everywhere inside the star, but its effect may be small (depending 
on the physical quantity under consideration) and may only be detectable at 
very high resolution. Fig.~\ref{fig:os_conv} shows that $E$
converges at second order, whereas $F$ converges at less
than second order but better than first order. This
shows that $F$ is more susceptible to propagation of the first
order 
behavior at the surface of the star, which can be anticipated by looking at the
shape of the profiles in Fig.~\ref{fig:rad_profile}: $E$ drops abruptly before reaching the surface, while $F$ increases 
monotonically up to the surface.  We see that convergence of $F$ deviates 
further from second order as the resolution is increased. This is consistent 
with the presence of a
first order term with small coefficient due to the discontinuity at 
the stellar surface: 
$F(\Delta,t)=F_{\rm exact}(t) + c_1(t)\Delta + c_2(t)\Delta^2 + O(\Delta^3)$. 
Here $F(\Delta,t)$ is the value of $F$ at time $t$ evolved with a grid size 
$\Delta$, $F_{\rm exact}(t)$ is the exact solution, and $c_1(t)$ and $c_2(t)$ 
are resolution-independent functions. The first-order term $c_1(t)\Delta$ 
results from the discontinuity. We expect that $c_1(t) \ll c_2(t)$ since 
we look at a point far away from the discontinuity. 
With a lower resolution, and hence a larger grid size
$\Delta$, the first-order term is not as significant relative to the
second order term because of the small coefficient. Once we
decrease $\Delta$ by increasing the resolution, the second-order
term diminishes as $\Delta^2$, while the first order term shrinks
as $\Delta$ only, making the first order term more
conspicuous. Similar behavior for E is not evident since E drops to a
very low value at the surface.
\label{os_tests}

\section{Conclusions}
\label{conclusions}
We have developed a code which can evolve the coupled
Einstein-Maxwell-MHD-Radiation equations in 3 + 1 dimensions. In the
implementation presented here 
this code is able to model the behavior of magnetized,
perfectly conducting, radiating fluids in dynamical spacetimes in
which the characteristic length scales of the system are longer than
the mean free path of the radiation and the opacity has a grey-body form.  
Our formalism allows us to evolve the radiation
fields using a HRSC scheme which is analogous to the
method we use for the hydrodynamic fields.  In this paper, we have tested the
shock-capturing capabilities of our code by simulating both continuous
and discontinuous one-dimensional radiating hydrodynamic waves.  We
have been successful in evolving highly relativistic, radiation-pressure
dominated, gas-pressure dominated, and Newtonian waves. We have
treated both stationary waves and boosted waves that propagate across
our computational grid in Minkowski spacetime. We have also
confirmed our ability to accurately capture the behavior of radiation
in a strong-field dynamical spacetime by simulating a thermal Oppenheimer-Snyder
collapse.  Our numerical results agree well with the analytic 
solutions. We perform convergence tests on our test problems and find the 
expected order of convergence in all cases.

We plan to use our radiation GRMHD code to study many interesting systems, 
and revisit some of the problems we have considered before, such as 
core-collapse supernovae, accretion onto a black hole, merging 
NSNSs and BHNSs, etc. By taking into account the effect of radiation, 
we hope to gain more insights and provide some answers to questions 
that are relevant to observations. For example, we can calculate the 
radiation luminosity, and study the radiation feedback to the dynamics 
of the systems.

\section{Acknowledgments}
\label{acknowledgements}
It is a pleasure to thank B. Stephens and Z. Etienne for useful
suggestions and discussions.  Some of these calculations were
performed at the National Center for Supercomputing Applications at the
University of Illinois at Urbana-Champaign (UIUC). 
This paper was supported in part by NSF
Grants PHY02-05155, PHY03-45151 and PHY06-50377 as well as 
NASA Grants NNG04GK54G
and NNX07AG96G at UIUC.

\appendix

\section{Limits}
\label{newtonian_appendix}
\subsection{Diffusion Approximation}

Using the form of $\Trad{\alpha}{\beta}$ given by Eq.~(\ref{R_def}) 
in the radiation moment equation
$\Trad{\alpha}{\beta}{}_{;\beta} = - G^{\alpha}$ gives
\begin{equation}
\label{rad_eqn_diff_approx}
\left(
\Erad u^{\alpha}u^{\beta} +
\Frad^{\alpha}u^{\beta} +
\Frad^{\beta}u^{\alpha} +
\Prad \proj{\alpha}{\beta}
\right)_{;\beta} = - G^{\alpha}
\end{equation}
where $\Prad=\Erad/3$.  In our formalism, we have already assumed near
isotropy, which implies that $\Frad^{\alpha}/\Erad \ll 1$.  We can
make the further approximation that the radiation flux 
$\Frad^{\alpha}$ can be neglected in the left hand side of Eq.~(\ref{rad_eqn_diff_approx}) to get
\begin{equation}
\left(
\Erad u^{\alpha}u^{\beta} +
\Prad h^{\alpha \beta}
\right)_{;\beta} = - G^{\alpha}
\end{equation}
Operating on both sides of the above equation with the projection
tensor $\projlow{\gamma}{\alpha}$, using Eq.~(\ref{eq:Ga}) for 
$G^{\alpha}$ and the fact that
$\Frad^\alpha u_{\alpha} = 0$, we get
\begin{eqnarray}
- \kappa \rho_0                                                              
\Frad^\gamma\ &=&
\projlow{\gamma}{\alpha} \left(
\Erad u^{\alpha}u^{\beta} +
\Prad h^{\alpha \beta}
\right)_{;\beta} \cr
&=& \projlow{\gamma}{\alpha} \left[
  \frac{4}{3} \Erad(u^{\alpha} u^{\beta})_{;\beta} 
 +\frac{1}{3} \Erad^{;\alpha}\right]\cr
&=& \frac{4}{3} \projlow{\gamma}{\alpha} \left(  
 \Erad a^{\alpha} 
 +\frac{1}{4} \Erad^{;\alpha}\right) \ ,
\end{eqnarray}
where $\kappa \equiv \kappa^a + \kappa^s$ is the total opacity, $a^{\alpha} 
\equiv u^{\alpha}{}_{;\beta}u^{\beta}$ is the 4-acceleration, and
where we have used the fact
that $\projlow{\gamma}{\alpha} u^{\alpha} = 0$ and 
$\Prad = \Erad/3$. Thus, we arrive at the expression for the radiation
flux in the diffusion approximation,
\begin{equation}
\mathbf{\Frad} = -  \frac{4}{3} \frac{1}{(\kappa^a + \kappa^s)
  \rho_0} \mathbf{\proj{}{}} \cdot \left(\frac{1}{4}\nabla \Erad + \mathbf{a}
  \Erad \right) \ . 
\end{equation}
This is the relativistic diffusion equation relating the radiation
flux $\Frad^{\alpha}$ to the local radiation energy density $\Erad$. 
If we further assume LTE, then $\Erad=a_RT^4$, in which case,
\begin{equation}
\mathbf{\Frad} = - \lambda_{th} \mathbf{\proj{}{}} \cdot (\nabla T + \mathbf{a}
T) \ ,
\end{equation}
where
\begin{equation}
\lambda_{th} = \frac{4}{3} \frac{a_R T^3}{(\kappa^a + \kappa^s) 
\rho_0} \ .
\end{equation}
This familiar result is in agreement with Eq.~(3.2) of
\cite{stu_rad_1}, and with Eq.~(2.5.28) of \cite{novikov_thorne}.  For
a further discussion of the simplification in
Eq.~(\ref{rad_eqn_diff_approx}) leading to the diffusion
approximation, see \cite{stu_rad_2}.
\label{diff_approx_appendix}
\subsection{Newtonian Limit}
It is instructive to consider the weak-field, slow-velocity (Newtonian) limit of the GR radiation 
hydrodynamic equations, and show that our equations reduce to the familiar 
expressions of Newtonian radiation hydrodynamics.
For simplicity, we set all large-scale electromagnetic fields to
zero (i.e. $\Tem{\mu}{\nu} = 0 = B^i$).

\subsubsection{Continuity equation}
From Eq. (\ref{continuity}), we have
\begin{equation}
(\rho_0 u^{\nu})_{;\nu}=0
\end{equation}
In Newtonian limit, the covariant derivative reduces to partial
derivative (in Cartesian coordinates), and the 4-vector 
$u^{\alpha}$ reduces to $u^{\alpha} \approx (1,v^i)$. Hence the continuity equation 
reduces to the familiar expression:
\begin{equation}
\label{newtonian_continuity}
\partial_t \rho_0 + \partial_j(\rho_0 v^j) =0 \ .
\end{equation}

\subsubsection{Euler equation}
From Eq.~(\ref{stress_energy_cons}), we have
\begin{equation}
\partial_t (\alpha \sqrt{\gamma} \Ttotallow{0}{i}) +
\partial_j(\alpha \sqrt{\gamma} \Ttotallow{j}{i}) = \frac{1}{2}\alpha
\sqrt{\gamma} \Ttotal{\alpha}{\beta} g_{\alpha \beta, i} \ .
\end{equation}
In Newtonian limit, the metric can be approximated by 
\begin{equation}
  ds^2 = -(1+2\Phi) dt^2 + (1-2\Phi) (dx^2+dy^2+dz^2) \ ,
\label{PN-metric}
\end{equation}
where $\Phi \ll 1$ is the Newtonian gravitational potential. 
Keeping only the lowest order terms, we obtain
\begin{equation}
\partial_t \Ttotallow{0}{i} +
\partial_j\Ttotallow{j}{i} = - \Ttotal{0}{0} \Phi_{,i} \ .
\end{equation}
Using Eq.~(\ref{stress_energy_total}) and assuming 
$\rho_0 \gg P$, $\rho_0 \gg \Erad$, and $\rho_0 v^i \gg \Frad^i$, 
we obtain
\begin{equation}
\label{newtonian_euler_conservative}
\partial_t (\rho_0 v_i) + \partial_j \left[\rho_0 v^j v_i + (P
+ \Prad) \delta^j{}_i\right] = -\rho_0 \partial_i \Phi \ .
\end{equation}
Combining Eq.~(\ref{newtonian_euler_conservative}) with the continuity
equation~(\ref{newtonian_continuity}) yields
\begin{equation}
\label{newtonian_euler}
\partial_t v_i + v^j \partial_j v_i = -\frac{1}{\rho_0} \partial_i(P + \Prad) - \partial_i \Phi  \ ,
\end{equation}
which is the familiar Newtonian Euler equation, allowing for gas plus
radiation pressure, ($P + \Prad$).
\subsubsection{Energy equation}
The energy equation in the Newtonian limit is derived by 
contracting $u_\mu$ with the equation $\nabla_\nu T^{\mu \nu}=0$. 
Using Eq.~(\ref{stress_energy_cons}) and the continuity 
equation $\nabla_\mu (\rho_0 u^\mu)=0$, we find, after some algebra, that
\begin{eqnarray}
 &&  u^\mu \nabla_\mu (\rho_0\epsilon +\Erad) 
 + (\rho_0\epsilon+\Erad+P+\Prad) \nabla_\mu u^\mu + \nabla_\mu F^\mu \cr 
 && \ \ \ \ \ + F^\mu a_{\mu} = 0 \ . 
\label{eq:energy_GR}
\end{eqnarray} 
where $a^{\mu}=u^\nu \nabla_\nu u^\mu$ is the 4-acceleration.
We note that $u^{\mu}a_{\mu}=0$ and $\Frad^{\mu}u_{\mu}=0$, which implies
that
\begin{equation}
\Frad^{\mu}a_{\mu} = \Frad^i a_i \left(1 + O(v^2)\right) \ .
\end{equation}
Furthermore, $\Frad^i \sim \Erad v^i$ and $a_i \sim v^j \partial_j v_i$,
so that
\begin{equation}
\Frad^{\mu}a_{\mu} \sim \Erad v^i v^j \partial_j v_i \ll E \partial_i
v^i \ ,
\end{equation}
so we may neglect $\Frad^{\mu}a_{\mu}$ in the Newtonian limit.
Also, $\partial_t \Frad^o \approx \partial_t(v_i \Frad^i) \sim
\partial_t (\Erad v^2) \ll \partial_t \Erad$ so we may neglect this
term as well.

Hence in the Newtonian limit, we obtain 
\begin{eqnarray}
  && \partial_t (\rho_0\epsilon +\Erad) 
  + v^i \partial_i (\rho_0\epsilon +\Erad) 
  + (\rho_0\epsilon+\Erad+P+\Prad) \partial_i v^i \cr
  && \ \ \ \ \ + \partial_i F^i = 0 \ .
\label{eq:energy_newtonian1}
\end{eqnarray}

This can be identified as the Newtonian energy equation for the
coupled fluid.

Equations~(\ref{eq:energy_GR}) and (\ref{eq:energy_newtonian1}) 
can be expressed in more familiar forms by introducing the total
(Lagrangian) time derivative comoving with the fluid:
\begin{equation}
  \frac{d}{d\tau} \equiv u^\mu \nabla_\mu \approx \frac{\partial}{\partial t} 
   + v^i \frac{\partial}{\partial x^i} \ .
\label{eq:dtau}
\end{equation}
The continuity equation $\nabla_\mu (\rho_0 u^\mu)=0$ gives 
$d \rho_0/d\tau = -\rho_0 \nabla_\mu u^\mu$. Hence, 
\begin{equation}
  \nabla_\mu u^\mu = -\frac{1}{\rho_0} \frac{d \rho_0}{d\tau} \ .
\label{eq:divu}
\end{equation} 
Combining Eqs.~(\ref{eq:energy_GR}), (\ref{eq:dtau}) and 
(\ref{eq:divu}) yields
\begin{eqnarray}
  && \frac{d}{d\tau} (\rho_0\epsilon+\Erad) - \frac{1}{\rho_0} 
(\rho_0\epsilon+\Erad+P+\Prad) \frac{d\rho_0}{d\tau} \cr
  && + \nabla_\mu F^\mu + F^\mu a_\mu = 0 \ ,
\label{eq:energy_GR2}
\end{eqnarray}
Dividing both sides of Eq.~(\ref{eq:energy_GR2}) by $\rho_0$ and
writing $E_{\rm tot} = \rho_0\epsilon+\Erad$ and $P_{\rm tot} = P+\Prad$,
we get
\begin{equation}
\label{general_first_law}
  \frac{d}{d\tau} \left( \frac{E_{\rm tot}}{\rho_0} \right) =
  -P_{\rm tot} \frac{d}{d\tau} \left( \frac{1}{\rho_0} \right)
  - \frac{1}{\rho_0} \nabla_\mu F^\mu - \frac{1}{\rho_0} F^\mu a_\mu \ .
\end{equation}
In Newtonian limit, $d/d \tau \rightarrow d/dt$, and we may neglect $\Frad^{\mu}a_{\mu}$ and $\partial_t
\Frad^0$ as explained above, and Eq.~(\ref{general_first_law})
reduces to 
\begin{equation}
\frac{d}{dt}\left(\frac{E_{\rm tot}}{\rho_0}\right) = -P_{\rm tot}
\frac{d}{dt}\left( \frac{1}{\rho_0} \right) - \frac{1}{\rho_0}
\nabla \cdot \mathbf{\Frad} \ ,
\end{equation}
which is the familiar first-law of thermodynamics in the case where
entropy is generated by radiation.

\subsubsection{Radiation equations}
From Eqs.~(\ref{rad_momentum}) and (\ref{rad_energy}), we get 
\begin{eqnarray}
\label{appendix_rad_momentum}
\partial_t \Srad{i} + \partial_j\left(\alpha \sqrt{\gamma} 
\Tradlow{j}{i} \right)
 &=& \alpha \sqrt{\gamma} \left(\frac{1}{2} \Trad{\alpha}{\beta}
 g_{\alpha \beta,i} -G_i\right) \ \ \ \ \  \\
\partial_t \taurad + \partial_i\left(\alpha^2 \sqrt{\gamma} \Trad{i}{0} \right)
 &=& \bar{s} -  \alpha^2 \sqrt{\gamma} G^0.
\label{appendix_rad_energy}
\end{eqnarray}
In Newtonian limit, 
\begin{eqnarray}
\Srad{i} &\approx& F_i \ , \\
\taurad &\approx & \Erad \ , \\
R^j{}_i &\approx & \Prad \delta^j{}_i \ , \\ 
R^{i0} &\approx & F^i.
\end{eqnarray}
Inserting these into (\ref{appendix_rad_momentum}) and 
(\ref{appendix_rad_energy}), and dropping higher
order terms, we obtain
\begin{eqnarray}
\label{newt_rad_1}
\partial_t \Frad_i+\partial_i \Prad &=& -G_i, \\ 
\label{newt_rad_2}
\partial_t \Erad + \partial_j F^j &=& -G^0, 
\end{eqnarray}
which agree with Eqs.~(94.2) and (94.3) in~\cite{mihalas}.

\section{Estimation of Characteristic Speeds}
\label{char_appendix}
In order to compute HLL fluxes, we must compute the maximum left-going
wave speed $c_-$ and maximum right-going wave speed $c_+$ on both
sides of the interface.  
We estimate the wave speed by computing the dispersion relation 
due to a small perturbation on a magnetized, radiating plasma of uniform 
$\rho_0$, $P$, $B^i$, $\Erad$ and $F^i$. In the comoving frame, 
$u^i=0$. We further choose our coordinate system so that 
the spacetime is locally Minkowski (i.e.\ $g_{\mu \nu} = \eta_{\mu \nu}$). 
To compute the dispersion relation, 
we consider a perturbation of the form
\begin{eqnarray}
\rho_0 &=& \bar{\rho}_0 + \delta \rho_0 e^{i(\ve{k}_{\mathrm{cm}} 
\cdot \ve{x} - \omega_{\mathrm{cm}} t)} \ ,\cr
P &=& \bar{P} + \delta P e^{i(\ve{k}_{\mathrm{cm}} \cdot \ve{x} 
- \omega_{\mathrm{cm}} t)} \ ,\cr
u^i &=& \delta u^i e^{i(\ve{k}_{\mathrm{cm}} \cdot \ve{x} 
- \omega_{\mathrm{cm}} t)} \ ,\cr 
B^i &=& \bar{B}^i + \delta B^i e^{i(\ve{k}_{\mathrm{cm}} \cdot \ve{x} 
- \omega_{\mathrm{cm}} t)} \ ,\cr
\Erad &=& \bar{\Erad} + \delta \Erad e^{i(\ve{k}_{\mathrm{cm}} \cdot \ve{x} 
- \omega_{\mathrm{cm}} t)} \ ,\cr
\Frad^i &=& \bar{\Frad}^i + \delta \Frad^i e^{i(\ve{k}_{\mathrm{cm}}\cdot\ve{x} 
- \omega_{\mathrm{cm}} t)} \ ,
\end{eqnarray}
where bar denotes the unperturbed quantity, and the subscript ``cm''
refers to comoving frame values.  Substituting these 12 expressions into
our 12 radiation-MHD equations
(\ref{rad_energy}), (\ref{rad_momentum}), 
(\ref{mhd_continuity}), (\ref{mhd_momentum})
and (\ref{mhd_energy}) and keeping terms linear in the perturbation, 
we obtain a matrix equation of the form $\ve{M} \ve{X}=0$.
Here $\ve{M}$ is a $12\times 12$ matrix, and 
$\ve{X}=(\delta \rho_0 \ \ \ \delta P \ \ \ \delta u^i \ \ \ 
\delta B^i \ \ \ \delta E \ \ \ \delta F^i)^t$, where superscript 
$t$ denotes the transpose. For simplicity, 
we drop the radiative source terms $G^\mu$ 
in deriving the dispersion relation (the curvature source terms 
vanish in Minkowski spacetime).
The dispersion relation is obtained by setting $\det(\ve{M})=0$, which
leads, after some algebra, to the following equation:
\begin{equation}
\omega_{\rm cm}^4 (\omega_{\rm cm}^2-k_{\rm cm}^2/3) 
[\omega_{\rm cm}^2-(\ve{k}_{\rm cm}\cdot \ve{v}_A)^2] 
Q(\omega_{\rm cm},\ve{k}_{\rm cm}) = 0 \ ,
\end{equation}
where 
\begin{eqnarray}
Q(\omega_{\rm cm},\ve{k}_{\rm cm}) &=& 
\omega_{\rm cm}^4 - [ k_{\rm cm}^2 c_m^2 + c_s^2 
(\ve{k}_{\rm cm} \cdot \ve{v}_A)^2 ] \omega_{\rm cm}^2 \cr 
 && + k_{\rm cm}^2 c_s^2 (\ve{k}_{\rm cm} \cdot \ve{v}_A)^2.
\end{eqnarray}
Here $c_s = \sqrt{\Gamma P / \rho_0 h}$ is the sound speed, 
$v_A=\sqrt{b^2/(\rho_0h+b^2)}$ is the Alfv\'en speed, 
and $c_m = \sqrt{v_A^2 + c_s^2 (1-v_A^2)}$. The solution 
$\omega_{\rm cm}=0$ corresponds to pure, stationary 
density perturbations, $\omega_{\rm cm}^2 = k_{\rm cm}^2/3$ 
is related to the propagation speed of the radiation flux for the nearly 
isotropic radiative diffusion, 
$\omega_{\rm cm}^2=(\ve{k}_{\rm cm}\cdot \ve{v}_A)^2$ corresponds 
to the Alfv\'en waves, and $Q(\omega_{\rm cm},\ve{k}_{\rm cm})=0$ 
corresponds to magnetosonic waves. As in~\cite{HARM}, we replace 
the dispersion relation $Q(\omega_{\rm cm},\ve{k}_{\rm cm})=0$ 
by $\omega_{\rm cm}^2 -c_m^2k_{\rm cm}^2 =0$ 
as it is more convenient when calculating 
the characteristic speed in the grid frame. As pointed out 
in~\cite{HARM}, this modified dispersion relation overestimates the 
maximum wave speed by a factor of $\leq 2$ in the comoving 
frame.

Since the HLL scheme only requires the information on the 
maximum and minimum characteristic speeds, we use the 
following dispersion relation to estimate the characteristic speeds:
\begin{eqnarray}
\label{eq:dispersion}
\frac{\omega_{\mathrm{cm}}}{k_{\mathrm{cm}}} = \left\{ \begin{array}{l}
\pm \sqrt{1/3}\\
\pm \sqrt{v_{\mathrm{A}}^2 + c_{\mathrm{s}}^2(1-v_{\mathrm{A}}^2)}
\end{array}\right. \ .
\end{eqnarray}

\section{Analytic Solutions for Radiation Tests in Minkowski Spacetime}
\label{analytic_shock_appendix}
\subsection{One-Dimensional waves in Minkowski Spacetime}

We begin by assuming a $\Gamma$-law EOS and write 
$\Gamma = 1+1/n$, where $n$ is the polytropic index.  Hence the EOS
(\ref{ideal_P}) becomes $\rho_0 \epsilon = n P$.

We consider a stationary, infinite fluid in Minkowski spacetime with
planar symmetry, hence we drop all time derivatives, all $y$ 
and $z$ components, and all $y$ and $z$ derivatives.  It follows from $\nabla_\mu (\rho_0 u^\mu)=0$, 
$\nabla_\nu T^{\mu \nu}=0$ and $\nabla_\nu R^{\mu \nu}=0$ that 
\begin{eqnarray}
 (\rho_0 u^x )_{,x} &=& 0 \ , \label{eq:1d-con} \\
 T^{0 x}{}_{,x}&=&0 \ , \\
 T^{x x}{}_{,x}&=&0 \ , \label{eq:txx} \\ 
 R^{0 x}{}_{,x}&=&-G^{0} \ , \label{eq:rox}\\
 R^{x x}{}_{,x}&=&-G^{x} \ . \label{eq:rxx}
\end{eqnarray}

It is convenient to define
\[\mathbf{P} = \left(
\begin{array}{c}
\rho_o\\
P\\
u^x\\
\Erad\\
\Frad^x
\end{array}\right),
\mathbf{U} = \left(
\begin{array}{c}
\rho_o u^x\\
T^{0 x} \\
T^{x x}\\
R^{0 x} \\
R^{x x}
\end{array}\right)\mathrm{\ and\ }
\mathbf{S} =\left(
\begin{array}{c}
0\\
0\\
0\\
 -G^0\\
 -G^x
\end{array}\right)\]
The system of ODEs in (\ref{eq:1d-con})--(\ref{eq:rxx}) becomes 
\[
\partial_x \mathbf{U}(\mathbf{P}   ) = \mathbf{S}(\mathbf{P}) \ .
\]

The first three equations [Eqs.~(\ref{eq:1d-con})--(\ref{eq:txx})] 
are readily integrated, giving three ``conserved quantities'' 
$U_1$, $U_2$, and $U_3$. In the presence of a shock, across which 
$\ve{P}$ is discontinuous, these ``conserved quantities'' 
give the Rankine-Hugoniot conditions, relating $\ve{P}$'s 
on both sides of the shock. The remaining two ODEs [Eqs.~(\ref{eq:rox}) 
and (\ref{eq:rxx})] can be integrated numerically using, for example, 
a fourth-order Runge-Kutta integrator. During the integration, we need to 
compute $\ve{P}$ from $\ve{U}$, which we outline as follows.
For simplicity, we consider the case without a large-scale electromagnetic 
field ($\Tem{\alpha}{\beta} = 0 = B^i$). Using Eq.~(\ref{stress_energy_total}) for 
$\Ttotal{\mu}{\nu}$, Eq.~(\ref{R_def}) for $R^{\mu \nu}$, and 
combining Eqs.~(\ref{eq:1d-con})--(\ref{eq:rxx}), we obtain
\begin{eqnarray} 
\label{1d_cont}
\rho_o u^x &=& U_1\\
\label{1d_T0x}
\left[ \rho_0 + (n+1)P \right] u^x u^0  = U_2-U_4 &\equiv& U_a\\
\label{1d_Txx}
\left[\rho_0 + (n+1)P  \right] (u^x)^2 + 
P = U_3-U_5 &\equiv& U_b\\
\label{1d_R0x}
\frac{4}{3}\Erad u^x u^0 + u^x \Frad^0 + u^0 \Frad^x &=& U_4\\
\label{1d_Rxx}
\frac{4}{3}\Erad (u^x)^2 + \frac{1}{3} \Erad +2 u^x \Frad^x  &=& U_5\ .
\end{eqnarray} 
Eliminating $\rho_0$ and $P$ from Eqs.~(\ref{1d_cont})--(\ref{1d_Txx}), 
we get an expression for $u^x$:
\begin{equation}
U_1u^0+(n+1)U_bu^0u^x-U_a[(u^0)^2+n(u^x)^2]=0 
\label{eq:ux}
\end{equation}
where $u^0 = \sqrt{1+(u^x)^2}$. Hence $u^x$ can be determined by 
solving the above algebraic equation. Having obtained $u^x$, the quantities
$\rho_0$ and $P$ are computed from 
\begin{eqnarray}
\rho_0&=&\frac{U_1}{u^x}\\
P&=&U_b-U_a\frac{u^x}{u^0} \ .
\end{eqnarray}
The values of $\Erad$ and $\Frad^x$ are obtained by solving the linear 
Eqs.~(\ref{1d_R0x}) and (\ref{1d_Rxx}). The result is 
\begin{eqnarray}
\Erad&=&\frac{\Delta_{\Erad}}{\Delta} \ ,\\
\Frad^x&=&\frac{\Delta_{\Frad}}{\Delta} \ ,
\end{eqnarray} 
where 
\begin{eqnarray}
\Delta&=&\frac{2}{3}u^0-\frac{1}{u^0}\\
\Delta_{\Erad}&=&2u^xU_4-u^0\left[\left(\frac{u^x}{u^0}\right)^2+1\right]U_5\\
\Delta_{\Frad}&=&\frac{4}{3}u^0u^x U_5-\left[\frac{4}{3}(u^x)^2+\frac{1}{3}\right]U_4  \ .
\end{eqnarray}

We note that Eq.~(\ref{eq:ux}) will, in general, have more than one 
real root.  Indeed, this must be the case in order for shocks to
exist.  In the absence of a shock, the appropriate root is chosen by continuity.

Our one-dimensional tests can be divided into two groups: fully
continuous configurations, and discontinuous configurations (i.e.\ 
shocks are present). In either case, we begin by specifying boundary 
conditions on the asymptotic left side ($x=-\infty$). In practice, 
we set up our computational domain with $x\in [-L,L]$. 
We specify $\rho_0$, $P$, and $u^x$ at $x=-L$, denoting 
them as $\rho_{0L}$, $P_L$ and $u_L^x$. 
We also impose that the radiation be in LTE with the gas at 
$x=-L$ [$\Erad_L = a_R T_L^4 = a_R m^4 (P_L/\rho_{0L})^4$] and set 
$\Frad_L^x = f_L$, where $f_L$ is a 
parameter chosen so that $\Frad_L^x/\Erad_L \ll 1$. 
Thus, we have specified 5
boundary conditions for our 5 ODE's. The values of $U_1$, $U_2$ and 
$U_3$, which are independent of $x$, are determined. 
If all quantities in the
configuration are continuous throughout the computational
domain, these boundary conditions at $x=-L$ are sufficient for us to
integrate Eqs.~(\ref{1d_R0x}) 
and (\ref{1d_Rxx}) from $x=-L$ to $x=L$. 

If however, we wish to determine a configuration that contains 
a discontinuity, which we set at $x=0$, we use a shooting method, 
described as follows.
In addition to the boundary conditions set at $x=-L$, we also
demand that the radiation be in LTE with the gas,
and that $\Frad^x = f_R$ at $x=L$. The parameter 
$f_R$ is chosen so that $\Frad^x/\Erad \ll 1$ at $x=L$.  
The constants $U_1$, $U_2$ and $U_3$ are fixed, giving the 
values of $\ve{P}$ at $x=-L$. Denote the values of $\rho_0$, 
$P$, $u^x$ and $\Erad$ at $x=L$ by $\rho_{0R}$, $P_R$, and 
$u^x_R$, and $\Erad_R$, respectively. The LTE condition at $x=L$ gives 
$\Erad_R = a_R m^4 (P_R/\rho_{0R})^4$. From the definition of 
$U_1$, $U_2$ and $U_3$, it is straightforward to show that 
\begin{eqnarray}
\label{PoRho}
\frac{P_R}{\rho_{0R}}=\left\{
\vphantom{\left(\frac{u_R^x}{u_R^0}\right)^2}
U_2-U_1 u^0_R+4\left[U_2(u^x_R)^2-U_3u_R^0u_R^x\right] \right.&&\cr
\left.\left. -f_R u_R^0\left[1-3\left(\frac{u_R^x}{u_R^0}\right)^2\right]\right\}
 \right/ \left[(n-3)U_1u_R^0\right] \ ,&&
\end{eqnarray}
\begin{eqnarray}
\label{u_root}
&&U_1u_R^0\left[1+(n+1)\frac{P_R}{\rho_{0R}}\right]+\frac{4}{3}
a_R m^4\left(\frac{P_R}{\rho_{0R}}\right)^4u_R^0u_R^x\cr
&&\indent +f_R\left[1+\left(\frac{u_R^x}{u_R^0}\right)^2\right]-U_2=0 \ ,
\end{eqnarray}
where $u_R^0 = \sqrt{1+(u_R^x)^2}$. 
Substituting Eq.~(\ref{PoRho}) into Eq.~(\ref{u_root}) gives 
an algebraic equation for $u_R^x$, which can be solved numerically. 
Obviously, $u_R^x = u_L^x$ is a solution, but we look for another 
solution in order to obtain a configuration containing a shock. 
Having determined $u_R^x$, the other quantities are computed as follows: 
\begin{eqnarray}
\rho_{0R}&=&U_1/u_R^x\\
P_R&=&\frac{P_R}{\rho_{0R}}\rho_{0R} \label{eq:Pr} \\
\Erad_R&=&a_R m^4\left(\frac{P_R}{\rho_{0R}}\right)^4 \label{eq:Er} \ ,
\end{eqnarray}
where $P_R / \rho_{0R}$ in above equations are computed from Eq.~(\ref{PoRho}). 
To generate the 1D shock configuration, we first specify 
$\rho_L$, $P_L$, $u_L^x$, and consider $f_L$ and $f_R$ as 
free parameters. The other quantities at $x=\pm L$ are fixed by the 
LTE condition at $x=\pm L$ and Eqs.~(\ref{PoRho})--(\ref{eq:Er}). 
We then integrate Eqs.~(\ref{eq:rox}) and (\ref{eq:rxx}) from both $x=\pm L$ to 
$x=0$. Since $U_1$, $U_2$ and $U_3$ are constants, the Rankine-Hugoniot junction conditions
are automatically satisfied at the shock front ($x=0$). Hence, we only have to
impose the junction conditions for the radiation variables, which are
the continuity of $R^{0x}$ and $R^{xx}$. In the Newtonian limit, these
conditions reduce to the continuity of $\Erad$ and $\Frad^x$ at the
shock front, but this is not the case in general.  In any case, we need two junction
conditions at the shock front, so a well-posed shooting problem requires
varying two boundary condition parameters until the solution satisfies
the two junction conditions at the shock front. We use $f_L$
and $f_R$ as such two parameters.

For a pure hydrodynamic (or MHD) shock, the profiles of $\ve{P}$ are 
constants on each side of the shock front. This is not the case for a radiating 
hydrodynamic shock, where $\ve{P}$ vary with $x$ and approach constants 
only in the asymptotic regions ($x\rightarrow \pm \infty$). This variation 
results from the radiative source terms $G^0$ and $G^x$, given by
Eq.~(\ref{eq:Ga}), which vanishes 
when the radiation and fluid are in strict LTE and the radiation flux
vanishes (i.e.\ in the asymptotic 
regions). The length scale over which the parameters vary between the two asymptotic regions 
is a few optical depths.  We need to choose 
$\kappa$ to ensure that $x=\pm L$ are in the asymptotic regions. 
In practice, one can estimate the optical depth 
$\tau=\int_{-L}^{L} \rho_0 \kappa dx \sim (\rho_{0L}+\rho_{0R})\kappa L$ 
and choose $\kappa$ so that $\tau \gg 1$. For our tests, we choose
$\tau \sim 10$.  Choosing 
a larger $\kappa$ does not change the profile (plotted against the 
rescaled coordinate $\kappa x$) significantly.

\subsection{Special Analytic Case}
In Newtonian limit, the solutions of (\ref{1d_cont})--(\ref{1d_Rxx}) 
can be written in analytic form under special conditions, as 
stated in~\cite{mihalas,zeldovich_raizer}. Here we briefly 
summarize this solution.

In Newtonian limit, Eqs.~(\ref{1d_cont})--(\ref{1d_Rxx}) become
\begin{eqnarray}
\rho_0 v &=& U_1\\
\left(\frac{1}{2} \rho_0 v^2 + \rho_0 \epsilon + P + \Erad + \Prad \right)
v +\Frad^x
&=& U_2\\
\rho_0 v^2 + P + \Prad &=& U_3 \ ,
\end{eqnarray}
while by dropping time derivatives in
Eqs.~(\ref{newt_rad_1})--(\ref{newt_rad_2}), we get
\begin{eqnarray}
\label{mihalas:diff_approx}
\frac{d \Frad^x}{d \tau} &=& 4 \pi B - \Erad\\
\label{mihalas:evolution_eqn}
\Frad^x &=& - \frac{1}{3} \frac{d E_{(r)}}{d \tau} \ ,
\end{eqnarray}
where $v=v^x$ and $\tau$ is the optical depth, given by 
$d \tau = \rho_0 \kappa d x$.
As in~\cite{mihalas,zeldovich_raizer}, 
we consider a strong shock propagating into cold gas, so that
the pressure and internal energy of the unshocked gas 
($x<0$)  can be neglected.  
We also assume that the gas is optically thick so that we may use the diffusion
approximation, and that it is sufficient to account for the radiation
flux, while neglecting radiation energy density and radiation
pressure. Under these assumptions, the above equations can be
rewritten as:
\begin{eqnarray} 
\rho_0 v &=& \rho_{0L} v_L\\
\rho_0 v^2 + P &=& \rho_{0L} v_L^2\\
\rho_0 v(\rho_0 \epsilon + v^2/2 ) + \Frad^x &=& \rho_{0L} v_L^2/2  \ .
\end{eqnarray}
Combining Eqs.~(\ref{mihalas:evolution_eqn}) and (\ref{mihalas:diff_approx})
gives 
\begin{equation} \label{rad:dif_eq}
\frac{d^2 \Frad^x}{d \tau^2} = 3 \Frad^x + 4 a_R T^3
\frac{dT}{d \tau}  \ .
\end{equation}
Solving these coupled equations using methods outlined 
in~\cite{mihalas,zeldovich_raizer} one arrives at:

Preshock Medium ($x<0$):
\begin{eqnarray}
\Frad^x &=& -\frac{1}{2\sqrt{3}} a_R T_R^4 e^{-\sqrt{3} |\tau|}\\
\Erad &=& \frac{1}{2} a_R T_R^4 e^{-\sqrt{3} |\tau|}
\end{eqnarray}

Postshock Medium ($x>0$):
\begin{eqnarray}
\Frad^x &=& -\frac{1}{2\sqrt{3}} a_R T_R^4 e^{-\sqrt{3} |\tau|}\\
\Erad &=& a_R T_R^4 (1- 1/2 e^{-\sqrt{3} |\tau|}) \ ,
\end{eqnarray}
Here $T_R$ is the asymptotic temperature in the postshock region 
and $\tau$ is measured from the shock front (i.e.\ $\tau(x) = \int_0^x 
\kappa \rho(x') dx'$).

\section{Thermal Oppenheimer-Snyder Solution}
\label{thermal_os_appendix}
Here we summarize the analytic solutions derived in~\cite{stu_rad_1}
for a strong field (black hole) dynamical scenario, 
used to compare with our 
numerical results in Section~\ref{os_tests}. 

The standard Oppenheimer-Snyder (OS) collapse solution for a
homogeneous dust ball was first
derived in~\cite{OS39}.  
In~\cite{stu_rad_1}, the collapsing sphere is subjected to radiation
and gas pressure perturbations which are assumed to be sufficiently 
small that the spacetime 
metric and density evolution are well-approximated 
by the OS solution. For this to be true, we require $P/\rho_0\ll M/R$ and
$\mathcal{P}/\rho_0=E/3\rho_0\ll M/R$ for a star with mass $M$ and
radius $R$. We are interested in the evolution of the radiation quantities 
$\Erad$ and $\Frad^i$ inside the star. 

Inside the star, the OS metric is given by the closed Friedmann line element
\begin{equation}
ds^2=-d\tau^2+a^2(\tau)(d\chi^2+\sin^2\chi d\Omega^2)
\label{os_metric}
\end{equation}
where $\tau$ is the proper time of a fluid element and $\chi$ is a
Lagrangian radial coordinate. The scale factor $a(\tau)$ is given in
parametric form according to
\begin{eqnarray}
a &=& \frac{1}{2}a_m(1+\cos\eta) \ , \label{eq:a_eta} \\
\tau &=& \frac{1}{2}a_m(\eta+\sin\eta) \ . \label{eq:tau_eta}
\end{eqnarray}
Here $\eta$ is the conformal time and
$a_m$ is a constant which is related to the initial areal radius $R_i$ of 
the star. (The subscript $i$ denotes initial values.) The radius is
given by 
\begin{equation}
R=\frac{1}{2}R_i(1+\cos\eta) \ . \label{eq:R_eta}
\end{equation} 
The exterior Schwarzschild line element is 
\begin{equation}
ds^2=-\left(1-\frac{2M}{r_s}\right)dt^2+\left(1-\frac{2M}{r_s}\right)^{-1}dr_s^2+r_s^2d\Omega^2
\label{sch_metric}
\end{equation}
where $r_s$ is the Schwarzschild (areal) radius. 
Matching the two metrics at $r_s=R$ gives 
\begin{equation}
a_m=\sqrt{\frac{R_i^3}{2M}} \ .
\end{equation}
Denote $\chi_0$ as the Lagrangian radial coordinate $\chi$ at the
stellar surface. It follows from Eqs.~(\ref{os_metric}), (\ref{sch_metric}), 
(\ref{eq:a_eta}) and (\ref{eq:R_eta}) that  
\begin{equation}
\sin\chi_0=\frac{R}{a}=\sqrt{\frac{2M}{R_i}} \ .
\end{equation}
In Friedmann comoving coordinates the density $\rho_0$ is always homogeneous and given by
\begin{equation}
\frac{\rho_0}{\rho_{0i}}=Q^{-3}
\end{equation}
where 
\begin{equation}
Q=\frac{a}{a_m}=\frac{1}{2}(1+\cos\eta) \ .
\end{equation}

An analytic solution for the $\Erad$ and $\Frad$ is derived 
in~\cite{stu_rad_1,stu_rad_2} by assuming (1) diffusion approximation, 
(2) that the radiation and fluid are in LTE ($E=a_RT^4$),   
and (3) that radiation pressure is much greater than gas pressure 
($\Prad \gg P$). We summarize the solution below. 

Define the radiation energy
density and flux ``corrected'' for adiabatic contraction as $E_c=Q^4E$ and $F_c=Q^4F$ respectively. Define a
time parameter $\tilde{\tau}$ as 
\begin{equation}
\tilde{\tau}=\frac{1}{4(\tau^a+\tau^s)}\sqrt{\frac{R_i}{8M}}\left(\frac{\sin{\chi_0}}{\chi_0}\right)^2\left(\eta+\frac{4}{3}\sin\eta+\frac{1}{6}\sin2\eta\right)
\end{equation}
where $\tau^a$ and $\tau^s$ are the initial absorption and scattering
optical depths respectively, related to the absorption and
scattering opacities $\kappa^a$ and $\kappa^s$ by
$\tau^a=\kappa^a\rho_{0i}R_i$ and $\tau^s=\kappa^s\rho_{0i}R_i$. (Note
that the analytic solution assumes $\kappa^a$ and $\kappa^s$ to be
constant throughout the collapse.)
Define also a normalized Lagrangian radius $z=\chi/\chi_0$, such that
$0\leq z\leq1$ within the star. Then the interior corrected energy
density is given by
\begin{eqnarray}
E_c(\tilde{\tau},z)&=&2E_i\left(\frac{\sin\chi_0}{\sin(\chi_0z)}\right)e^{\chi_0^2\tilde{\tau}}\sum^\infty_{n=1}\left[(-1)^{n+1}\vphantom{\frac{\pi}{\chi_0^2}}
\right.\nonumber\\
&&\left.e^{-n^2\pi^2\tilde{\tau}}\sin(n\pi
z)\frac{n\pi}{n^2\pi^2-\chi_0^2}\right] 
\label{Ec}
\end{eqnarray}
where $E_i$ is the initial value of $E$, assumed to be 
constant throughout the star. Note that in the
derivation of Eq.~(\ref{Ec}) the ``zero temperature'' boundary condition
($E=0$ at the stellar surface) has been used.

The diffusion approximation gives the expression for the corrected radiation
flux $F_c$ in terms of the gradient of $E_c$:
\begin{eqnarray}
F_c(\tilde{\tau},z)&=&-\frac{Q^2}{3}\left(\frac{\sin\chi_0}{\chi_0}\right)
\left(\frac{1}{\tau^a+\tau^s}\right)\frac{\partial
  E_c}{\partial z} \\ 
&=&\frac{2}{3}\frac{E_iQ^2}{\tau^a+\tau^s}\frac{1}{\chi_0}\left(\frac{\sin\chi_0}{\sin(\chi_0z)}\right)^2e^{\chi_0^2\tilde{\tau}}\sum^\infty_{n=1}\left\{(-1)^{n+1}\vphantom{\frac{\pi}{\chi_0^2}}\right.\nonumber\\
&&e^{-n^2\pi^2\tilde{\tau}}[\chi_0\sin(n\pi
  z)\cos(\chi_0 z)\nonumber\\
&&\left.-n\pi\sin(\chi_0 z)\cos(n\pi z)]\frac{n\pi}{n^2\pi^2-\chi_0^2}\right\} 
\ .
\end{eqnarray}

Finally, an analytic expression for the ideal gas pressure $P=\rho_0T/m$ is obtained from
the LTE condition $E=a_RT^4=a_Rm^4\left(\frac{P}{\rho_0}\right)^4$. Hence
\begin{equation}
P=\rho_0\left(\frac{E}{a_Rm^4}\right)^{1/4}\ ,
\end{equation}
where $\rho_0$ and $E$ are given by the analytic solutions above.

A comparison between the analytic solution for thermal OS collapse and
the exact solution of the Boltzmann equation of radiative transfer for
the same problem is given in \cite{stu_rad_2}.

\bibliography{radiation}
\end{document}